\documentclass[12pt]{article}
\usepackage{a4}
\usepackage{amsmath}
\usepackage{amssymb}
\usepackage[dvips]{graphicx}
\usepackage{subfigure}
\def\beq{\begin{equation}}
\def\eeq{\end{equation}}
\def\beqa{\begin{eqnarray}}
\def\eeqa{\end{eqnarray}}
\def\n{\nonumber \\}

\def\ep{\epsilon}

\newcommand{\bel}{\begin{equation}\label}

\newcommand {\bc}{\begin{center}}
\newcommand {\ec}{\end{center}}
\newcommand {\tr}{{\rm tr}\,}

\newcommand {\CTr}{{\cal T}r\,}

\def\dag{\dagger}
\def\vs5{\vspace*{5mm}}
\def\vs1{\vspace*{1cm}}
\def\vs2{\vspace*{2cm}}
\def\hs5{\vspace*{5mm}}
\def\hs1{\hspace*{1cm}}
\def\hs2{\hspace*{2cm}}
\def\vs50{\vspace*{50mm}}
\def\vs20{\vspace*{20mm}}

\begin{document}
%%%%%%%%%%%%%%%%%%%%%%%%%%%%%%%%%%%%%%%%%%%%%%%%
 
\begin{flushright}
{SAGA-HE-230}\\
{KEK-TH-1110}
\end{flushright}
\vskip 0.5 truecm
%%%%%%%%%%%%%%%%%%%%%%%%%%%%%%%%%%%%%%%%%%%%%%%%
 
%%%%%%%%%%%%%%%%%%%%%%%%%%%%%%%%%%%%%%%%%%%%%%%%
 
\begin{center}
{\Large{\bf Ginsparg-Wilson Dirac operator 
in the monopole backgrounds on the fuzzy 2-sphere}}
\vskip 1.0cm
 
{\large Hajime Aoki$^{a}$\footnote{e-mail
 address: haoki@cc.saga-u.ac.jp},
 Satoshi Iso$^{b}$\footnote{e-mail
 address: satoshi.iso@kek.jp}
and 
 Toshiharu Maeda$^{a}$\footnote{e-mail
 address: maeda@th.phys.saga-u.ac.jp}\\
}
\vskip 0.5cm

$^a${\it Department of Physics, Saga University, Saga 840-8502,
Japan  }\\
  
$^b${\it Institute of Particle and Nuclear Studies, \\
High Energy Accelerator Research Organization (KEK)\\
Tsukuba 305-0801, Japan}

\end{center}
 
\vskip 1cm
\begin{center}
\begin{bf}
Abstract
\end{bf}
\end{center}
In the previous papers, we studied the 't Hooft-Polyakov (TP) 
monopole configurations
in the $U(2)$ gauge theory on the fuzzy 2-sphere,
and showed that they have nonzero topological charges
in the formalism based on the Ginsparg-Wilson (GW) relation.
In this paper, 
we will show an index theorem in the TP monopole background,
which is defined in the projected space,
and provide a meaning of the projection operator. 
%we will provide an interpretation of the projection operator 
%which was introduced to define the topological charge for the TP monopoles.
We also extend the index theorem to general configurations which
do not satisfy the equation of motion,
and show that the configuration space can be classified into 
the topological sectors.
We further calculate 
the spectrum of the GW Dirac operator in the TP monopole backgrounds,
and consider the index theorem in these cases.
%We also obtain the exact form of the chiral zero-modes.

%%%%%%%%%%%%%%%%%%%%%%%%%%%%%%%%%%%%%%%%%%%%%%%% 
\newpage
\setcounter{footnote}{0}
\section{Introduction}
\setcounter{equation}{0}

Matrix models are a promising candidate to 
formulate the superstring theory nonperturbatively \cite{Banks:1996vh,IKKT}, 
%In IIB matrix model, 
%which is the 
%dimensionally reduced model of the 10-dimensional super 
%Yang-Mills theory, 
where
both spacetime and matter are 
described in terms of matrices, and noncommutative (NC) 
geometries\cite{Connes} naturally appear\cite{CDS,NCMM,Seiberg:1999vs}. 
One of the important subjects of the matrix model 
is a construction of 
configurations with nontrivial indices in finite NC geometries, 
%It is not only interesting 
%from the mathematical point of view 
%but necessary from the physical requirement. 
since compactifications of extra dimensions
with nontrivial indices can realize 
chiral gauge theories in our spacetime. 
Topologically nontrivial configurations in NC geometries were constructed 
by using algebraic K-theory and projective modules 
\cite{non-trivial_config,non_chi,Valtancoli:2001gx,
Steinacker:2003sd,Karabali:2001te,Carow-Watamura:2004ct}.
%However the relation to indices of Dirac operators 
%was not clear in these formulations. 

%The formulation of NC geometries
%in Connes' prescription is based on the spectral
%triple (${\cal A}$,${\cal H}$,${\cal D}$), where a chirality
%operator and a Dirac operator which anti-commute are
%introduced\cite{Connes}. 
%Since NC geometries on compact manifolds have only 
%finite degrees of freedom, 
%a more suitable framework to discuss the problems mentioned 
%above will be to modify the Connes' spectral triple so that 
In order to see their relation to indices of Dirac operators,
a suitable framework will be the one where
the chirality operator and the Dirac operator satisfy the
Ginsparg-Wilson (GW) relation\cite{GinspargWilson},
since NC geometries on some compact manifolds have only 
finite degrees of freedom.
The GW relation has been developed in the lattice gauge theory.
Its explicit construction was given by
the overlap Dirac operator \cite{Neuberger}
and the perfect action \cite{Hasenfratzindex}.
The exact chiral symmetry\cite{Luscher,Nieder} and 
the index theorem\cite{Hasenfratzindex,Luscher} 
at a finite cutoff can be realized 
due to the GW relation.

In ref.\cite{AIN2}, we have provided 
a general prescription to construct 
a GW Dirac operator
%chirality operators and a Dirac operator satisfying the GW relation 
with coupling to nonvanishing gauge field backgrounds
on general finite NC geometries.
As a concrete example 
we considered the fuzzy 2-sphere\cite{Madore}\footnote{
The GW Dirac operator on the fuzzy 2-sphere for vanishing gauge field 
was constructed earlier in \cite{balagovi}.}.
%%%%%%%%%%%%%%
%On the fuzzy 2-sphere two types of Dirac operators, 
%$D_{\rm WW}$\cite{Carow-Watamura:1996wg} and
%$D_{\rm GKP}$\cite{Grosse:1994ed,IKTW}, had been constructed. 
%$D_{\rm WW}$ has doublers and the correct chiral anomaly 
%cannot be reproduced. 
%On the other hand, $D_{\rm GKP}$ breaks 
%chiral symmetry at finite matrix size, 
%and the chiral structures are not 
%clear,
%though the chiral anomaly 
%can be reproduced correctly in the commutative 
%limit\cite{chiral_anomaly,non_chi,chiral_anomaly2,AIN1}.
%%%%%%%%%%
%Hence the formalism based on the GW relations is 
%more suitable for studying chiral structures on the fuzzy 2-sphere.
%
%We constructed GW Dirac operator in 
%general gauge field configurations in \cite{AIN2}.
Owing to the GW relation, an index theorem can be proved 
even for finite NC geometries.
We have defined a topological charge,
and showed that it takes only integer values, 
and becomes the Chern character in the commutative limit
\cite{AIN2,Ydri:2002nt,Balachandran:2003ay,AIN3}\footnote{
The GW relation was implemented also on the NC 
torus by using the Neuberger's overlap Dirac 
operator in \cite{Nishimura:2001dq}.  
In \cite{Iso:2002jc},
this GW Dirac operator was also derived from the general prescription 
\cite{AIN2} and
the correct chiral anomaly was 
reproduced  by using a topological 
method in \cite{Fujiwara:2002xh}. 
The correct parity anomaly was reproduced in 
\cite{Nishimura:2002hw}.}.

We then constructed the 't Hooft-Polyakov (TP) monopole configuration
as a topologically nontrivial configuration
\cite{Balachandran:2003ay,AIN3}.
We showed that this configuration 
is a NC analogue of the commutative TP monopole 
by explicitly studying the form of the configuration.
We then redefined the topological charge 
by inserting a projection operator,
and showed that it reproduces the correct form of 
the topological charge in the commutative limit.
%namely, the magnetic flux for the unbroken $U(1)$ component.
We also showed that  the topological charge takes the appropriate values
for the TP monopole configurations. 
Furthermore, in \cite{AIMN},
we presented a mechanism for the 
dynamical generation of a nontrivial index,
by showing that 
the TP monopole configurations 
are stabler than the topologically trivial sector
in the Yang-Mills-Chern-Simons matrix model
\cite{Myers:1999ps,IKTW}
\footnote{
The stability of these configurations are also
investigated  in papers 
\cite{IKTW,Bal:2001cs,Valtancoli:2002rx,Imai:2003vr,Azuma:2004zq,Castro-Villarreal:2004vh}.}.

In this paper,
we will prove an index theorem in the TP monopole backgrounds.
The TP monopole configuration breaks
the $SU(2)$ gauge symmetry down to $U(1)$,
and the matter field in the fundamental representation of 
the $SU(2)$ gauge group
%in this background
has two components, corresponding to $+1/2$ and $-1/2$ electric charges
of the unbroken $U(1)$ gauge group.
Since these two components cancel the index and the chiral anomaly,
we need to pick up one of them. 
The index with a projection operator to pick up one of the two components is 
shown to give the above topological charge 
introduced in ref. \cite{Balachandran:2003ay,AIN3,AIMN}.
%We therefore consider an index theorem with a projection operator
%to pick up $+1/2$ or $-1/2$ electric charge. 
%This projection operator is the same as the one
%introduced in ref. \cite{Balachandran:2003ay,AIN3,AIMN}.
%and mentioned in the previous paragraph.
%Then
%and it gives a physical meaning of the projection operator.

The index theorem can be extended to more general 
configurations which do not satisfy the equation of motion.
By modifying the chirality operators and the GW Dirac operator
in general gauge field configurations on the fuzzy 2-sphere,
we propose a topological charge classifying 
configurations in spontaneously symmetry broken gauge theories.
This topological charge is shown to become the 't Hooft's 
topological charge in the commutative limit.
Since the $U(2)$ gauge theory on the fuzzy sphere is generally broken down to 
$U(1) \times U(1)$ gauge theory through Higgs mechanism, 
this generalization shows that the configuration space
of gauge fields on the fuzzy sphere can be classified 
into the topological sectors. 
We also discuss the validity of this classification by introducing the 
concept of admissibility condition, 
which was developed to investigate 
the topological structure in the lattice gauge theory.

We also calculate 
the spectrum of the GW Dirac operator in the TP monopole backgrounds\footnote{
While we are preparing the manuscript, we are informed of 
the related work \cite{ISTT} before their publication,
where they study the monopole harmonics on 
commutative sphere and fuzzy sphere.
},
and confirm the index theorem in these cases.
The spectrum was also provided in \cite{Balachandran:2003ay}.
%Since the configurations of the TP monopoles in \cite{AIN3}
%are differnt from \cite{Balachandran:2003ay} except for $m =0,1$,
%we obtain differet eigenvalues.
We study the spectrum in more details and
obtain the explicit forms of the eigenstates by using the GW relation.
The largest eigenvalue states are shown to
play an important role in the index theorem.

In section 2 we briefly review 
how to define the GW Dirac operator on the fuzzy 2-sphere
and how to construct the TP monopole configurations.
We then show the index theorem in the TP monopole backgrounds, and
give an interpretation of the projection operator.
We further extend the index theorem to general configurations.
In section 3, we calculate the spectrum for the 
GW Dirac operator in the TP monopole backgrounds,
and in section 4, we obtain the forms of the chiral zero-modes.
Section 5 is devoted to conclusions and discussions.
In appendix A, we calculate the spectrum for another type of Dirac operator
$D_{\rm GKP}$, which is given in (\ref{DGKP}).
In appendix B, we abtain the spectrum for the Dirac operator in
the commutative theory.
  
%%%%%%%%%%%%%%%%%%%%%%%%%%%%%%%%%%%%%%%%%%%%%%%% 

\section{Index theorem in spontaneously-symmetry-broken gauge theory
on fuzzy 2-sphere}
\setcounter{equation}{0}

\subsection{Review on fuzzy 2-sphere}
\label{sec:f2s}

NC coordinates of the
fuzzy 2-sphere are described by
\begin{equation}
x_i =\alpha L_i, \label{xi}
\end{equation}
where $\alpha$ is the NC parameter,
and $L_i$'s are $n$-dimensional irreducible
representation matrices of $SU(2)$ algebra: 
\begin{eqnarray}
[L_i,L_j]&=&i\epsilon_{ijk}L_k . \label{su2}
\end{eqnarray}
Then we have the following relation,
\begin{eqnarray}
(x_i)^2&=&\alpha^2 \frac{n^2-1}{4}{\bold 1}_n 
\equiv \rho^2 {\bold 1}_n,
\end{eqnarray}
where 
$\rho=\alpha \sqrt{\frac{n^2-1}{4}}$ 
expresses the radius of the fuzzy 2-sphere.
The commutative limit can be taken by $\alpha \to0, n \to \infty$
with $\rho$ fixed.
 
Any wave functions on the fuzzy 2-sphere are mapped to 
$n \times n$ matrices. 
We can expand them in terms of NC
analogues of the spherical harmonics $\hat{Y}_{lm}$, 
which are traceless symmetric products of
the NC coordinates, and has an upper bound for 
the angular momentum $l$ as $l \le n-1$.
Derivatives along the Killing vectors of a function $M(\Omega)$ 
on the 2-sphere 
are written as the adjoint operator of $L_i$ on the corresponding 
matrix $\hat{M}$:
\begin{equation}
{{\cal L}_i}M(\Omega)=-i\epsilon_{ijk}x_j \partial_k M(\Omega)
\leftrightarrow
{\tilde L}_i \hat{M}= [L_i, \hat{M}] =(L_i^L-L_i^R)\hat{M}.
\end{equation}
Here the superscript L (R) in $L_i$ means that this
operator acts from the left (right) on the matrix $\hat{M}$. 
An integral of functions is given by a trace
of matrices:
\begin{equation}
\int \frac{d \Omega}{4 \pi}M(\Omega)
\leftrightarrow \frac{1}{n} \tr[\hat{M}]. 
\end{equation}

Two types of Dirac operators, 
$D_{\rm WW}$\cite{Carow-Watamura:1996wg} and
$D_{\rm GKP}$\cite{Grosse:1994ed,IKTW}, were constructed. 
$D_{\rm WW}$ has doublers and the correct chiral anomaly 
cannot be reproduced. 
On the other hand, $D_{\rm GKP}$ breaks 
chiral symmetry at finite matrix size, 
and the chiral structures are not 
clear,
though the chiral anomaly 
can be reproduced correctly in the commutative 
limit\cite{chiral_anomaly,non_chi,chiral_anomaly2,AIN1}.
We here review $D_{\rm GKP}$.  
The fermionic action is defined as
\begin{eqnarray}
S_{\rm GKP}&=& \tr[\bar\Psi D_{\rm GKP} \Psi], \\
D_{\rm GKP}&=& \sigma_i ({\tilde L}_i + \rho a_i ) +1,
\label{DGKP}
\end{eqnarray}
where $\sigma_i$'s are Pauli matrices. 
The gauge field $a_i$ of $U(k)$ gauge group
and the fermionic field $\Psi$ in the fundamental representation of the gauge group
are expressed by $nk\times nk$ and $nk\times n$ matrices, respectively.  
This action is invariant under the gauge transformation:
\begin{eqnarray}
\Psi &\rightarrow& U \Psi, \n
\bar\Psi &\rightarrow& \bar\Psi U^\dagger , \n
a_i &\rightarrow& U a_i U^\dag +\frac{1}{\rho} (U L_i U^\dag-L_i),
\label{gaugeTra}
\end{eqnarray}
since a combination,
which sometimes called a covariant coordinate,
\begin{equation}
A_i \equiv L_i+\rho a_i
\label{defAi}
\end{equation}
transforms covariantly as
\begin{equation}
A_i \rightarrow U A_i U^\dagger. \label{gaugetr_A}
\end{equation}
 
In the commutative limit, 
the Dirac operator (\ref{DGKP}) becomes 
\begin{equation}
D_{\rm GKP} \rightarrow 
D_{\rm com}=\sigma_i ({\cal L}_i + \rho a_i ) +1,
\end{equation}
which is the ordinary Dirac operator on the commutative 2-sphere. 
The gauge fields $a_i$'s in 3-dimensional space can be 
decomposed into the tangential components on the 
2-sphere $a_i'$ and the normal component $\phi$ as
\begin{eqnarray}
&&\left\{
\begin{array}{lll}
a_i'&=& \epsilon_{ijk}n_j a_k, \\
\phi&=&n_i a_i,
\end{array}
\right. \label{decomposeto}\\
&\Leftrightarrow& a_i = -\epsilon_{ijk}n_j a_k' + n_i \phi,
\label{decomposefrom}
\end{eqnarray}
where $n_i = x_i/\rho$ is a unit vector.
The normal component $\phi$ is a scalar field on the 2-sphere.
Then, the Dirac operator $D_{\rm com}$ and $D_{\rm GKP}$
have a coupling to the scalar field. 

%Due to the noncommutativity of the coordinates, 
%$D_{\rm GKP}$ does not anti-commute with
%the chirality operator.
%Then, 
%by carefully evaluating this nonzero anticommutation relation,
%the chiral anomaly 
%can be reproduced correctly 
%\cite{chiral_anomaly,non_chi,chiral_anomaly2,AIN1}.
%However, chiral structures are not transparent 
%in this formulation,
%and we will define another Dirac operator
%suitable for these issues.

\subsection{GW Dirac operator and index theorem}
\label{sec:GWindex}

In order to discuss the chiral structures, 
a Dirac operator satisfying the GW relation is more suitable. 
Ref \cite{AIN2} provided a general prescription to
define GW Dirac operator in arbitrary gauge 
field backgrounds on general finite NC geometries.
We first define two chirality operators:
\begin{eqnarray}
\Gamma^R &=& a\left(\sigma_i L_i^R -\frac{1}{2}\right)
, \label{gammaR}\\
\hat\Gamma&=&\frac{H}{\sqrt{H^2}} \label{gammahat},
\end{eqnarray}
where
\begin{equation}
H=a\left(\sigma_i A_i +\frac{1}{2}\right), 
\end{equation}
$A_i$ is defined in (\ref{defAi}), and
\begin{equation}
a=\frac{2}{n}
\end{equation}
is introduced as a NC analogue of a lattice-spacing.
These chirality operators satisfy
\begin{equation}
(\Gamma^R)^\dagger=\Gamma^R, \
(\hat\Gamma)^\dagger=\hat\Gamma, \
(\Gamma^R)^2=(\hat\Gamma)^2=1.
\end{equation}
In the commutative limit, both $\Gamma^R$ and $\hat\Gamma$
become the chirality operator on the commutative 2-sphere,
$\gamma = n_i \sigma_i$.

We next define the GW Dirac operator as
\begin{equation}
D_{\rm GW} = -a^{-1}\Gamma^R (1- \Gamma^R \hat{\Gamma}).
\label{defDGW}
\end{equation}
Then the action
\begin{equation}
S_{\rm GW}= \tr [\bar\Psi D_{\rm GW} \Psi]
\end{equation}
is invariant under the gauge transformation
(\ref{gaugeTra}).
In the commutative limit, $D_{\rm GW}$ becomes 
\begin{equation}
D_{\rm GW} \rightarrow 
D'_{\rm com}=\sigma_i ({\cal L}_i + \rho P_{ij} a_j ) +1,
\label{DGWcom}
\end{equation}
where $P_{ij}=\delta_{ij}-n_i n_j$ 
is the projector to the tangential directions 
on the sphere. Thus this Dirac operator $D'_{\rm com}$ is nothing but the Dirac operator on the commutative 2-sphere
without coupling to the scalar field .

By the definition (\ref{defDGW}),
the GW relation  
\begin{equation}
\Gamma^R D_{\rm GW}+D_{\rm GW} \hat{\Gamma}=0 
\label{GWrelation}
\end{equation}
is satisfied. 
Then the following index theorem is satisfied:
\begin{equation}
{\rm{index}}(D_{\rm GW})\equiv (n_+ - n_-)=\frac{1}{2}
{\cal T}r(\Gamma^R +\hat{\Gamma}), 
\label{indexth}
\end{equation}
where
$n_{\pm}$ is the number of zero-modes of $D_{\rm GW}$
with positive or negative chirality (for either $\Gamma^R$
or $\hat{\Gamma}$),
and ${\cal T}r$ represents a trace 
over the space of matrices and over the spinor index.
(See \cite{AIN2,AIN3} for a proof.)

The rhs of (\ref{indexth}) has the following properties. 
Firstly, it takes only integer values since both $\Gamma^R$ 
and $\hat{\Gamma}$ have a form of sign operator 
by the definitions (\ref{gammaR}), (\ref{gammahat}). 
Secondly, it becomes the topological charge on the 2-sphere, 
the Chern character,
in the commutative limit \cite{AIN2,AIN3}. 
Finally, it takes nonzero values for topologically nontrivial configurations
if we slightly modify the definition of it,
which we will see in the next subsections.

\subsection{Monopole configurations}
\label{sec:monopole}

As topologically nontrivial configurations
in the $U(2)$ gauge theory on the fuzzy 2-sphere,
the following monopole configurations were constructed \cite{Balachandran:2003ay,AIN3}:
\begin{equation}
A_i=
\begin{pmatrix}
 L_i^{(n+m)} & \cr
& L_i^{(n-m)} \cr
\end{pmatrix} 
\label{LnpmLnmm}, 
\end{equation}
where $A_i$ is defined in (\ref{defAi}),
and $L_i^{(n\pm m)}$ are $(n\pm m)$ dimensional 
irreducible representations of $SU(2)$ algebra\footnote{
Since (\ref{LnpmLnmm}) with $-m$ is unitary equivalent to 
the one with $m$,
we will restrict $m \ge 0$ without loss of generality
in this paper.
}.
The total matrix size is $N=2n$.
The $m=0$ case corresponds to two coincident fuzzy 2-spheres,
whose effective action is given by the $U(2)$ gauge theory
on the fuzzy 2-sphere.
The cases with  general $m$ correspond to 
two fuzzy 2-spheres 
which share the same center but have the different radii.
For $m \ll n$,
as we will see below, 
they correspond
to the monopole configurations with magnetic charge $-m$,
where the gauge group $U(2)$ is spontaneously broken down
to $U(1) \times U(1)$.
%(see below (\ref{TCunbrokenprocom})).
%The $m=n$ case corresponds to a single fuzzy 2-sphere,
%whose effective action is the $U(1)$ gauge theory
%on the fuzzy 2-sphere.

For the $m=1$ case,
(\ref{LnpmLnmm}) is unitary and gauge equivalent to the following configuration:
\begin{equation}
U A_i U^\dagger= L_i^{(n)} \otimes {\bold 1}_2 +
{\bold 1}_{n} \otimes \frac{\tau_i}{2} ,
\label{AiLtau}
\end{equation}
as is easily seen from the $SU(2)$ algebra.
From (\ref{defAi}),
the first and the second terms represent 
the coordinate of the NC space and the configuration 
of the $U(2)$ gauge field respectively.
Then, the gauge field is given by 
\begin{equation}
a_i=\frac{1}{\rho}{\bold 1}_{n} \otimes \frac{\tau_i}{2}.
\label{monopoleconfig}
\end{equation}
By taking the commutative limit of (\ref{monopoleconfig}), 
and decomposing it into the normal and the tangential components
of the sphere as in (\ref{decomposeto}), 
it becomes 
\beqa
a_i^{\prime a} &=& \frac{1}{\rho} \ep_{ija} n_j, \\
\phi^a &=& \frac{1}{\rho} n_a,
\eeqa
which is precisely the TP monopole configuration
\cite{AIN3}.

\subsection{Index theorem in the monopole backgrounds}
\label{sec:Index}
The index (\ref{indexth}) turns out to vanish
for the TP monopole configurations (\ref{LnpmLnmm}) with general $m$.
The reason is as follows.
TP monopole configuration breaks
the $SU(2)$ gauge symmetry down to $U(1)$. 
Then the fermionic field in the fundamental representation 
which couples to the TP monopole background
contains two components, corresponding to $+1/2$ and $-1/2$ electric charge
of the unbroken $U(1)$ gauge group.
Hence, these two components cancel the index.
We thus need to pick up one of the two components 
in order to obtain nonzero index and chiral anomaly.

As in (\ref{indexth}), the following equality is satisfied 
also in the projected space\footnote{
Note that the projection operator $P^{(n \pm m)}$ is written by the
Casimir operator $(A_i)^2$ as in (\ref{P1Ai}).
Then, we can see from the definition (\ref{gammaR}) and (\ref{gammahat})
that
$[P^{(n \pm m)}, \Gamma^R]=[P^{(n \pm m)}, \hat\Gamma]=0$.
Also, from (\ref{defDGW}), we can see
$[P^{(n \pm m)}, D_{\rm GW}]=0$.
}:
\begin{equation}
{\rm{index}}(P^{(n \pm m)}D_{\rm GW})=\frac{1}{2}
{\cal T}r[P^{(n \pm m)}(\Gamma^R +\hat{\Gamma})], 
\label{indextheorem}
\end{equation}
where $P^{(n \pm m)}$ is the projection operator to pick up
the Hilbert space for
the $n \pm m$ dimensional representation in (\ref{LnpmLnmm}).
%\footnote{
%This projection operator is same as the one 
%introduced in \cite{Balachandran:2003ay,AIN3}.
%In this paper, we will provide its meaning 
%by switching to another representation and investigating (\ref{Tother}).
%}
That is, the projection operator picks up 
one of the two fuzzy 2-spheres.
%The equality (\ref{indextheorem}) is satisfied
%and the rhs takes only integer values  
%since the proof given in \cite{AIN2,AIN3} is also valid 
%even in the projected space.
%From (\ref{LnpmLnmm}),  
The projection operator is written as
\begin{eqnarray}
P^{(n \pm m)}
&=& \frac{(A_i)^2-\frac{1}{4}[(n \mp m)^2-1]} 
{\frac{1}{4}[(n \pm m)^2-1]-\frac{1}{4}[(n \mp m)^2-1]} 
\label{P1Ai} \\
&=& \frac{1}{2}(1\pm T),\label{P1phi}
\end{eqnarray}
where 
\begin{eqnarray}
T &=& \frac{2}{nm}\left(A_i^2 - \frac{n^2+m^2-1}{4}\right) \label{defT}\\
  &=& \begin{pmatrix}
 {\bold 1}_{(n+m)} & \cr
& -{\bold 1}_{(n-m)} \cr
\end{pmatrix}.\label{1npm1nmp}
\end{eqnarray}

On the other hand, in the representation (\ref{defAi}), 
 (\ref{defT}) becomes 
\begin{equation}
T  =\frac{2}{nm}\left(\rho \{ L_i , a_i \}+ \rho^2 a_i^2-\frac{m^2}{4}\right).
\label{Tother}
\end{equation}
In the commutative limit, it becomes
$\frac{2\rho}{m} \phi$ when $m\ll n$,
where $\phi$ is the scalar field 
defined in (\ref{decomposeto}).
Then, $T$ is proportional to the scalar field.
Also, $T$ is normalized as $T^2={\bf 1}_{2n}$, 
which can be seen from (\ref{1npm1nmp}).
Therefore, $T$ is the generator for 
the unbroken $U(1)$ gauge group in the TP monopole.
Then, the eigenstate of $T$ with  eigenvalue $\pm 1$
corresponds to the fermionic state with $\pm 1/2$
electric charge of the unbroken $U(1)$ gauge group
\footnote{
Strictly speaking, 
since $T \sim \frac{2\rho}{m} \phi$,
the eigenstate of $T$ with eigenvalue $\pm 1$
corresponds to electric charge $\pm 1/2$ ($\mp 1/2$)
for $m >0$ ($m <0$).
Anyway, we restrict $m \ge 0$ in this paper.
}.
Thus the projection operator $P^{(n \pm m)}$ picks up  
$\pm 1/2$ electric charge component.

Then, the index in the projected space (\ref{indextheorem}) 
corresponds to the index for each electric charge component,
which is precisely what we needed to define
as we mentioned at the beginning of this subsection.
For the configurations (\ref{LnpmLnmm}),
we can see \cite{Balachandran:2003ay,AIN3}
\begin{equation}
\frac{1}{2}{\cal T}r[P^{(n \pm m)}(\Gamma^R +\hat{\Gamma})]
=\mp m.
\label{TPpmm}
\end{equation}
%by the straightforward algebraic calculations
%\cite{Balachandran:2003ay,AIN3}.
We thus obtain nonzero index $\mp m$ for $\pm 1/2$ electric charge component.
Without the projection operator, contributions from $+1/2$ and $-1/2$ 
charges cancel the index.

In the commutative limit, 
\beqa
\frac{1}{2}{\cal T}r[P^{(n \pm m)}(\Gamma^R +\hat{\Gamma})]
&=&\frac{1}{2}{\cal T}r[\frac{1}{2}(1 \pm T)(\Gamma^R +\hat{\Gamma})] \n
&=&\pm \frac{1}{2}{\cal T}r[\frac{1}{2}T (\Gamma^R +\hat{\Gamma})] \n
&\longrightarrow& 
\pm \frac{\rho^2}{8\pi}\int_{S^2} d\Omega \epsilon_{ijk}
n_i \phi'^a F_{jk}^a , 
\label{TCunbrokenprocom}
\eeqa
where $\phi'^a$ is a scalar field 
normalized as $\sum_a (\phi'^a)^2=1$.
%and can be defined as $T =\phi'^a \tau^a$ in the NC theory.
$F_{jk}=F_{jk}^a \tau^a/2$ 
is the field strength defined as
$F_{jk}= \partial_j a_k'-\partial_k a_j'-i[a_j',a_k']$.
This is 
%nothing but 
%the Chern character with the projection 
%to the component $\phi'^a$,
%namely,
the magnetic charge for the unbroken $U(1)$ component
in the TP monopole configuration,
which is nothing but the topological charge for the TP monopole 
configuration\footnote{
The topological charge should have an additional term as 
the second term in 
(\ref{TCforTP}).
However the additional term vanishes for the TP monopole configurations.
}.
Compared with (\ref{TPpmm}), 
the topological charge defined by inserting $\phi'$ as in 
(\ref{TCunbrokenprocom}) turns out to be $-|m|$ for the configurations
(\ref{LnpmLnmm}),
as in \cite{AIMN}.

Finally we see the gauge symmetry breaking 
in the configurations (\ref{LnpmLnmm}) with $m \ge 1$:
$U(2) \to U(1) \times U(1)$.
There are two ways to look at the unbroken gauge symmetries\footnote
{We gave the similar argument in \cite{AIMN} for $m=1$ case.
We here generalize it to $m\ge 1$.}:
\begin{enumerate}
\item Each sphere in (\ref{LnpmLnmm}) 
has unbroken $U(1)$ symmetry, and totally (\ref{LnpmLnmm}) 
has $U(1)\times U(1)$ symmetry. 
The generator for each $U(1)$ is written as\beq
\begin{pmatrix}
 {\bold 1}_{(n+m)} & \cr
& 0 \cr
\end{pmatrix}, \,\,
\begin{pmatrix}
 0 & \cr
& {\bold 1}_{(n-m)} \cr
\end{pmatrix}.
\label{unbrogene}
\eeq
%This geometrical description is clear in the representation (\ref{LnpmLnmm}).
\item $U(2) \simeq  SU(2) \times U(1)$, and the $SU(2)$ 
breaks down to $U(1)$ in the TP monopole 
configuration.
We can rearrange the generators (\ref{unbrogene})
as  
\beq
{\bold 1}_{2n}, \,\, T,
\eeq
where $T$ is given by (\ref{1npm1nmp}).
On the other hand, in the representation (\ref{Tother}),
$T$ is identified as noncommutative generalization of 
the unbroken $U(1)$ generator for 
the commutative TP monopole configurations.
\end{enumerate}
The above two ways of looking at the unbroken gauge symmetry
are equivalent since 
the representations (\ref{LnpmLnmm}) and  (\ref{defAi})
are unitary and gauge equivalent.

\subsection{Extension to general configurations}
In the previous subsection, 
we have considered the index theorem (\ref{indextheorem})
 for the monopole background configurations (\ref{LnpmLnmm}),
which satisfy the equation of motion. 
We now extend it to general configurations which do not
necessarily satisfy the equation of motion.
The only assumption in the following is that the 
$U(2)$ gauge symmetry is spontaneously broken to $U(1) \times U(1)$
through the Higgs mechanism, i.e. a nonzero value of 
the scalar field. 
Under this assumption,
the gauge configuration space on the fuzzy 2-sphere can be classified 
into the topological sectors. 

We first generalize the definition of the electric charge operator $T$ to
\beq
T' = \frac{(A_i)^2-\frac{n^2-1}{4}}{\sqrt{[(A_i)^2-\frac{n^2-1}{4}]^2}}.
\label{defTgen}
\eeq
This definition is valid for general configurations $A_i$
unless the denominator has zero-modes.
It satisfies 
\beq
(T')^\dag =T', \ (T')^2 = 1,
\eeq
and its eigenvalue takes $1$ or $-1$. 
The commutative limit of $T'$ becomes the normalized scalar field as
\beq
T' \to 2 \phi'=2\phi'^a \frac{\tau^a}{2}.
\eeq
For the configurations (\ref{LnpmLnmm}),
$T'$ reduces to the previous one (\ref{1npm1nmp}).

We next define modified chirality operators as
\beqa
\Gamma' &=& \frac{\{T', \Gamma^R \}}{2} = T' \Gamma^R, \\
\hat\Gamma' &=& \frac{\{T', \hat\Gamma \}}{\sqrt{\{T', \hat\Gamma \}^2}}.
\eeqa
These chirality operators are weighted by the electric charge operator $T'$ 
 but they still satisfy the usual relations: 
\beq
(\Gamma')^\dag = \Gamma', \ (\hat\Gamma')^\dag =\hat\Gamma', \
(\Gamma')^2 = (\hat\Gamma')^2 =1.
\eeq

We then define a modified GW Dirac operator as
\beq
D'_{\rm GW} = -a^{-1} \Gamma' (1 - \Gamma' \hat\Gamma').
\label{defDGWgen}
\eeq
This Dirac operator is also  weighted by the electric charge operator $T'$, 
which avoids the cancellation between the contributions from 
$\pm1/2$ electric charge components when we consider its index.
In the commutative limit, we obtain
\beq
D'_{\rm GW} \to \frac{1}{2} \{2\phi', D'_{\rm com} \}.
\eeq
In the $\phi'^a = (0,0,1)$ gauge, it becomes
\beq
\tau^3 (\sigma_i {\cal L}_i +1 + \rho \sigma_i P_{ij} a_j^3 \frac{\tau^3}{2}).
\eeq
Inside of the parenthesis is precisely the Dirac operator with coupling to 
the unbroken $U(1)$ gauge field. 

From the definition (\ref{defDGWgen}), this Dirac operator satisfies the GW relation
\beq
\Gamma' D'_{\rm GW} + D'_{\rm GW} \hat\Gamma' =0
\eeq
and thus the index theorem
\beq
\frac{1}{2} {\rm index}(D'_{\rm GW}) = \frac{1}{4} {\cal T}r [\Gamma' + \hat\Gamma']
\label{indextheogen}
\eeq
can be proved similarly to the ordinary case.
In the commutative limit, the rhs turns out to become
\beq
\frac{1}{4} {\cal T}r [\Gamma' + \hat\Gamma']
\to
\frac{\rho^2}{8\pi}\int_{S^2} d\Omega \epsilon_{ijk}
n_i \Bigl( \phi'^a F_{jk}^a 
- \epsilon_{abc} \phi'^a (D_j \phi'^b) (D_k \phi'^c) \Bigr),
\label{TCforTP}
\eeq
which is precisely the topological charge for
the configurations with unbroken $U(1)$ gauge symmetry
\cite{'tHooft:1974qc}.

For the configurations (\ref{LnpmLnmm}),
$T'$ commutes with with $\hat\Gamma$.
Then we obtain $\hat\Gamma' = T'\hat\Gamma$.
Then the above index reduces to the previous one;
\beq
\frac{1}{4} {\cal T}r [\Gamma' + \hat\Gamma']
=\frac{1}{2} {\cal T}r [\frac{1}{2} T(\Gamma^R + \hat\Gamma)].
\eeq
Hence the index theorem (\ref{indextheogen})
gives a natural generalization for general 
configurations which are not restricted to 
the special solutions such as the TP monopoles.

Finally we consider a condition for
gauge configurations that
the topological charge (\ref{indextheogen})
can be well-defined. 

%and we provide a sufficient condition on it.
In the commutative limit, we obtain
\beq
(A_i)^2-\frac{n^2-1}{4} 
\to \rho n \phi(x)
= \rho n \phi(x)^a \frac{\tau^a}{2}.
\eeq
Then,
\beq
\bigl[ (A_i)^2-\frac{n^2-1}{4}\bigr]^2
\to 
\frac{(\rho n)^2}{4} [ \sum_a(\phi^a(x))^2 {\bf 1}_2 + {\cal O}(1/n) ].
\eeq
In order to define the topological sectors 
by using the unbroken $U(1)$ gauge symmetry
in the commutative theory,
the scalar field should take non-vanishing values 
on arbitrary points on the sphere,
namely,
$
\rho^2 \sum_a(\phi^a(x))^2 \sim {\cal O}(1)
$
for all $x$.
Otherwise we could not define the unbroken $U(1)$ direction,
nor could we define the topological charge (\ref{TCforTP}).
This condition corresponds to 
\beq
\bigl[ (A_i)^2-\frac{n^2-1}{4}\bigr]^2
\sim {\cal O}(n^2),
\label{extadmcond}
\eeq
which means that all of the eigenvalues are of the order of $n^2$.
Smaller eigenvalues may invalidate the definition of topology, 
while larger eigenvalues may change the structure of space
and violate the assumption that we define the gauge theory 
on the fuzzy 2-sphere.
The upper bound on the eigenvalues corresponds to 
the admissibility condition,
which was introduced to assure the topological structure in
the lattice gauge theory \cite{Luscher:1981zq,Hernandez:1998et,Luscher:1998kn}. 
The condition (\ref{extadmcond}) has also the lower bound,
then it gives an extension of the admissibility condition.

More detailed analysis of this subsection will be reported in a
separated paper \cite{aokiiso}.

\section{Spectrum of the GW Dirac operator}\label{spectDGW}
\setcounter{equation}{0}
In this section, we will calculate the spectrum 
for the Ginsparg-Wilson Dirac operator
(\ref{defDGW})
in the monopole backgrounds (\ref{LnpmLnmm}):
\begin{equation}
D_{\rm GW}^m = 
               \begin{pmatrix}
               \frac{n}{n+m}(\sigma_i L_i^{(n+m)} +\frac{1}{2}) & \cr
               & \frac{n}{n-m}(\sigma_i L_i^{(n-m)} +\frac{1}{2}) \cr
               \end{pmatrix} -(\sigma_i L_i^R -\frac{1}{2}).
\label{gwdm}
\end{equation}

We define the total angular momentum operator
\begin{eqnarray}
M_i&=&L_i+a_i-L_i^R + \frac{\sigma_i}{2}  \label{defMLa}\\
   &=&A_i-L_i^R + \frac{\sigma_i}{2}  \label{defMA}\\
   &=&\begin{pmatrix}
      L_i^{(n+m)} & \cr
        & L_i^{(n-m)}  \cr
      \end{pmatrix} -L_i^R + \frac{\sigma_i}{2}.
\eeqa      
%T&=&\frac{2}{nm}\biggl( A_i^2 -\frac{n^2+m^2-1}{4}\biggr) \\
%&=&\begin{pmatrix}
%      {\bold 1}_{(n+m)} & \cr
%        & -{\bold 1}_{(n-m)}  \cr
%      \end{pmatrix} \\
%\Gamma^R &=& \frac{2}{n}\left(L^R\cdot\sigma -\frac{1}{2}\right),
%\end{eqnarray}
%where $M_i$ is a .
We also consider
the electric charge operator $T$ defined in (\ref{defT}),
and the chirality operator $\Gamma^R$ defined in (\ref{gammaR}).
Since $M_i, T, \Gamma^R$ commute with one another,
%\beqa
%\left[ M_i \ ,\  T \right] &=& 0, \\
%\left[ M_i \  ,\  \Gamma^R \right] &=& 0, \\
%\left[ T \  ,\  \Gamma^R \right] &=& 0, 
%\eeqa
we can consider the simultaneous eigenstates for these operators as\footnote{
Since $M_i$, $T$, and $\hat\Gamma$ also commute with one another,
we can consider the simultaneous eigenstates for them,
and follow the same calculations that will be performed in 
this paper by using $M_i$, $T$, and $\Gamma^R$.}
\begin{eqnarray}
M_i^2|J,J_3,\delta , \nu \rangle &=&J(J+1) \ |J,J_3,\delta , \nu \rangle, 
\label{sesmm}\\
M_3|J,J_3,\delta , \nu \rangle &=&J_3 \ |J,J_3,\delta , \nu \rangle, \\
T|J,J_3,\delta , \nu \rangle &=&\delta \ |J,J_3,\delta , \nu \rangle, \\
\Gamma^R|J,J_3,\delta , \nu \rangle &=&\nu \ |J,J_3,\delta , \nu \rangle.
\label{sesgammar}
\end{eqnarray}

We here note that  
the state with $\delta =\pm 1$ is
the state with spin $L \pm m/2$ for the operator $A_i$,
where $L$ is taken to be $\frac{n-1}{2}$.
The state with $\nu =\pm 1$ is 
the state with spin $L \mp 1/2$ for the operator $-L_i^R + \frac{\sigma_i}{2}$.
Thus, from (\ref{defMA}),
$J$ takes the values given in Table \ref{Mmeigens}.
Here we assumed $m > 0$.
\begin{table}[t]
\caption{Values of $J$ in each $\delta, \nu$ sector.
We assume $m>0$ here.}
\label{Mmeigens}
\begin{center}
\begin{tabular}{|c@{\quad\vrule width0.8pt\quad}ccccc|} 
\hline
                         & $\delta \ (T)$     & $-$     & $-$     & $+$     & $+$     \\
$J$                      & $\nu \ (\Gamma^R)$ & $+$     & $-$     & $+$     & $-$     \\
\noalign{\hrule height 0.8pt}
$\frac{m-1}{2}$          &                    & $\circ$ &         &         & $\circ$ \\
$\frac{m+1}{2}$          &                    & $\circ$ & $\circ$ & $\circ$ & $\circ$ \\
$\frac{m+3}{2}$          &                    & $\circ$ & $\circ$ & $\circ$ & $\circ$ \\
\rotatebox{90}{$\cdots$} &                
                                              &\rotatebox{90}{$\cdots$} 
                                                        &\rotatebox{90}{$\cdots$} 
                                                                  &\rotatebox{90}{$\cdots$} 
                                                                            &\rotatebox{90}{$\cdots$} \\
$2L-\frac{m+1}{2}$       &                    & $\circ$ & $\circ$ & $\circ$ & $\circ$ \\
$2L-\frac{m-1}{2}$       &                    &         & $\circ$ & $\circ$ & $\circ$ \\
%$                $       &                    &         &         & $\circ$ & $\circ$ \\
\rotatebox{90}{$\cdots$} &                    &         & 
                                                                  &\rotatebox{90}{$\cdots$} 
                                                                            &\rotatebox{90}{$\cdots$} \\
$2L+\frac{m-1}{2}$       &                    &         &         & $\circ$ & $\circ$ \\
$2L+\frac{m+1}{2}$       &                    &         &         &         & $\circ$ \\
\hline
\end{tabular}
\end{center}
\end{table}
Since the total number of these eigenstates is
\begin{eqnarray}
&&2\left[2\cdot\frac{m-1}{2}+1\right]
+4\left[\sum_{J=\frac{m+1}{2}}^{2L-\frac{m+1}{2}}(2J+1)\right] 
+3\left[2\Bigl(2L-\frac{m-1}{2}\Bigr)+1\right] \n 
&&+2\left[\sum_{J=2L-\frac{m}{2}+\frac{3}{2}}^{2L+\frac{m-1}{2}}(2J+1)\right]
+\left[2\Bigl(2L+\frac{m+1}{2}\Bigr)+1\right] \n
%&=&4(2L+1)^2\\
&=&4n^2,
\end{eqnarray}
these eigenstates exhaust the complete set of the Hilbert space.
Note that the spectrum starts from the nonzero
lowest spin, $J=\frac{m-1}{2}$, 
as in the case of the monopole harmonics.
Note also that, for the lowest spin states with $J=\frac{m-1}{2}$,
and for the highest spin states with
$(J,\delta)=(2L-\frac{m-1}{2}, -1)$,
$(2L+\frac{m+1}{2}, +1)$,
only $\nu=+1$ or $\nu=-1$ exists for each $\delta$,
while for the other states, 
both  $\nu=+1$ and $\nu=-1$ exist for each $\delta$\footnote{
The unbalance between $\nu=+1$ and $\nu=-1$ in the total spectrum 
is consistent with ${\cal T}r (\Gamma^R)=-4n$.
The unbalance between $\delta=+1$ and $\delta=-1$ in the total spectrum 
is consistent with ${\cal T}r (T)=4nm$.
}
\footnote{
We can also consider the simultaneous eigenstates for $M_i$, $T$, and $\hat\Gamma$,
and make the same table as 
Table \ref{Mmeigens}.
For the lowest spin states, 
which is shown to be zero-modes of $D_{\rm GW}$,
$\pm$ eigenvalue for $\Gamma^R$
corresponds to $\pm$ eigenvalue for $\hat\Gamma$,
as can be seen from the definition of $D_{\rm GW}$
(\ref{defDGW}).
For the highest spin states, $\pm$ eigenvalue for $\Gamma^R$
corresponds to $\mp$ eigenvalue for $\hat\Gamma$,
as can be seen from the GW relation
(\ref{GWrelation}).
}.

By straightforward calculations,
square of $D_{\rm GW}^m$ becomes
\begin{equation}
(D_{\rm GW}^{m})^2=\frac{n}{n+ mT}\left[ M_i^2-\frac{m^2-1}{4}\right],
\end{equation}
and $(D_{\rm GW}^{m})^2$ commutes with $M_i, T, \Gamma^R$.
%\beqa
%\left[ M_i \  ,\  (D_{\rm GW}^{m})^2\right] &=& 0, \\
%\left[ T \  ,\  (D_{\rm GW}^{m})^2\right] &=& 0, \\
%\left[ \Gamma^R \  ,\  (D_{\rm GW}^{m})^2\right] &=& 0.  
%\eeqa
We thus obtain the spectrum for $(D_{\rm GW}^m)^2$ as follows:
\beq
(D_{\rm GW}^{m})^2|J,J_3,\delta , \nu \rangle =
\frac{n}{n+ m\delta}\left[ J(J+1)-\frac{m^2-1}{4}\right]
 \ |J,J_3,\delta , \nu \rangle.
\eeq
Note that the states with the lowest spin $J=\frac{m-1}{2}$ correspond to 
the zero-modes for the  $D_{\rm GW}^{m}$.
%and $J > \frac{m-1}{2}$ to nonzero-modes.

We can also show
\beqa
\left[ M_i \  ,\  D_{\rm GW}^{m} \right] &=& 0, \\
\left[ T \  ,\  D_{\rm GW}^{m} \right] &=& 0,\\
\left[ \Gamma^R \  ,\  D_{\rm GW}^{m} \right] &\ne& 0. 
\eeqa
Therefore, linear combinations over $\nu$ for each $J,J_3,\delta $,
$\sum_{\nu}c_{\nu}|J,J_3,\delta , \nu \rangle$, 
give eigenstates for the Dirac operator $D_{\rm GW}^m$ as
\begin{eqnarray}
D_{\rm GW}^{m}
\left[\sum_{\nu}c_{\nu}|J,J_3,\delta , \nu \rangle\right] 
&=&\pm \sqrt{\frac{n}{n+ m\delta}\left[ J(J+1)-\frac{m^2-1}{4}\right]}
\left[\sum_{\nu}c_{\nu}|J,J_3,\delta , \nu \rangle\right] \n
&&
\label{esgwdm}  
\end{eqnarray}

By the Ginsparg-Wilson relation,
\beqa
\Gamma^R D_{\rm GW}^{m}+ D_{\rm GW}^{m} \hat{\Gamma}&=&0, 
\label{gwrel1m}\\
D_{\rm GW}^{m} \Gamma^R+\hat{\Gamma} D_{\rm GW}^{m}&=&0,
\label{gwrel2m}
\eeqa
if $\sum_{\nu}c_{\nu}|J,J_3,\delta , \nu \rangle$ 
is an eigenstate for the $D_{\rm GW}^{m}$ with eigenvalue $\alpha$,\\
$(\Gamma^R+\hat{\Gamma})\sum_{\nu}c_{\nu}|J,J_3,\delta , \nu \rangle$
is an eigenstate with eigenvalue $-\alpha$.
%\beqa
%D_{\rm GW}^{m}
%\left[\sum_{\nu}c_{\nu}|J,J_3,\delta , \nu \rangle\right] 
%&=&
%\alpha\left[\sum_{\nu}c_{\nu}|J,J_3,\delta , \nu \rangle\right],\\
%D_{\rm GW}^{m}
%\left[(\Gamma^R+\hat{\Gamma})\sum_{\nu}c_{\nu}|J,J_3,\delta , \nu \rangle\right] 
%&=&
%-\alpha\left[(\Gamma^R+\hat{\Gamma})\sum_{\nu}c_{\nu}|J,J_3,\delta , \nu \rangle\right],
%\eeqa
%where
%\beq
%\alpha=
%\pm \sqrt{\frac{n}{n+ m\delta}\left[ J(J+1)-\frac{m^2-1}{4}\right]}.
%\label{dgweigenvaluem}
%\eeq
%
From (\ref{defDGW}),
we see $\Gamma^R+\hat{\Gamma}=aD_{\rm GW}^{m}+2\Gamma^R$.
Then,
\beq
(\Gamma^R+\hat{\Gamma})\sum_{\nu}c_{\nu}|J,J_3,\delta , \nu \rangle
=(a\alpha+2)c_{1}|J,J_3,\delta , 1 \rangle
+(a\alpha-2)c_{-1}|J,J_3,\delta , -1 \rangle .
\label{GRpGhes}
\eeq
For the states with the highest spin,
($J = 2L-\frac{m-1}{2}$, $\delta=-1$)
and ($J = 2L+\frac{m+1}{2}$, $\delta=1$) in Table \ref{Mmeigens},
only $\nu = -1$  exists. 
Hence $(\Gamma^R+\hat{\Gamma})|J,J_3,\delta , \nu =-1\rangle$
must vanish.
From (\ref{GRpGhes}) we obtain $a\alpha=2$.
Thus only positive eigenvalues for $D_{\rm GW}^{m}$
exist for the highest spin states.
For the other states, 
%due to (\ref{dgweigenvaluem}), 
since $\alpha=
\pm \sqrt{\frac{n}{n+ m\delta}\left[ J(J+1)-\frac{m^2-1}{4}\right]}$,
we see
$-2 < a\alpha < 2$.
Since $(\Gamma^R+\hat{\Gamma})\sum_{\nu}c_{\nu}|J,J_3,\delta , \nu \rangle$
does not vanish,
both positive and negative eigenvalues exist. 
%except for the highest or the lowest spin states. 
We illustrate the spectrum for the case with $n=10$ and  $m=0,1,2,3,4$
in Figure \ref{fig:spe}.
\begin{figure}[hp]
\begin{center}
\includegraphics[height=18cm]{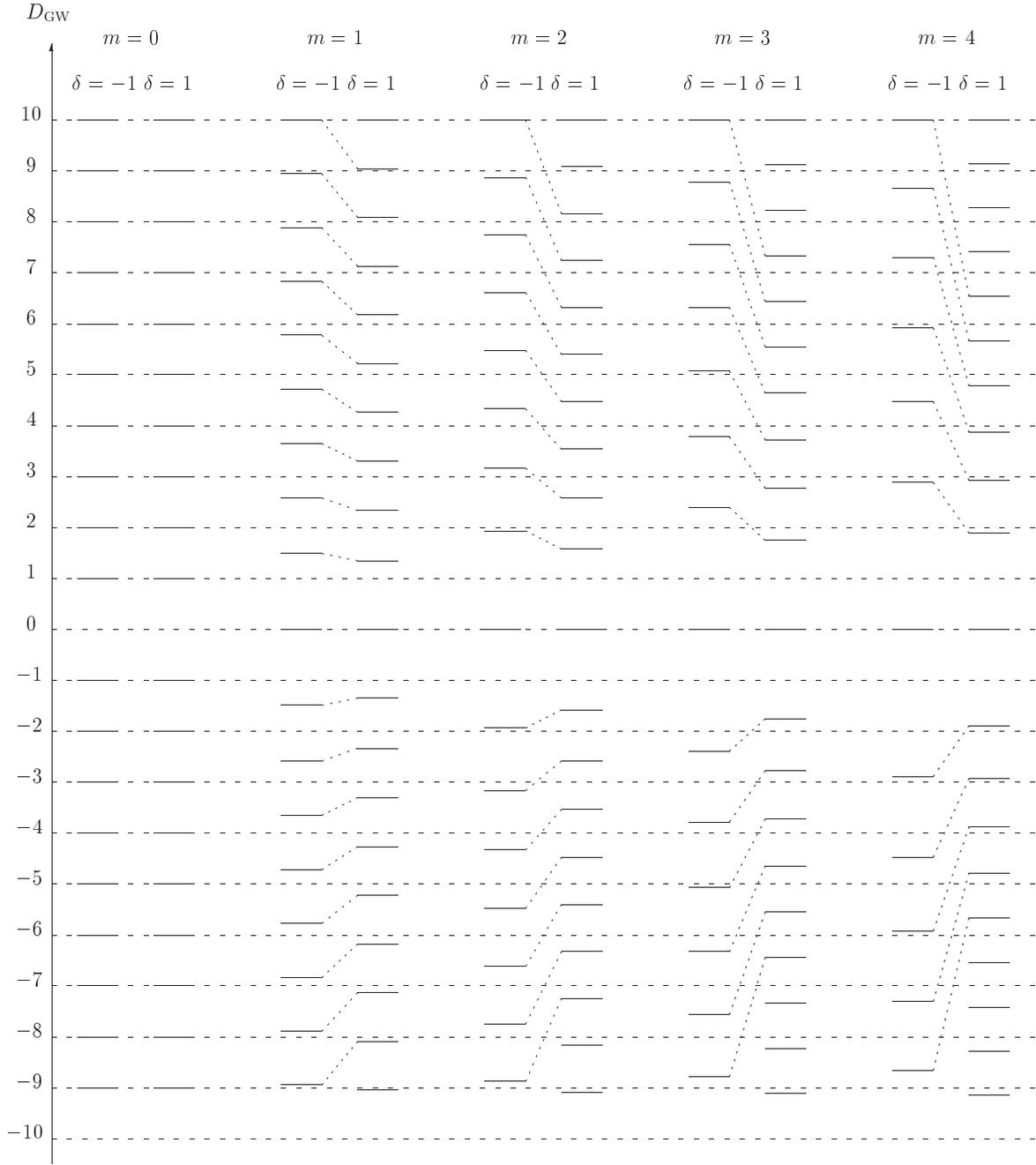}
\end{center}
\caption{Spectrum for GW Dirac operator for the cases of $n=10$ and 
$m=0,1,2,3,4$.
Doted line connects the states with the same value of $J$.}
\label{fig:spe}
\end{figure}

Next, we determine the coefficients $c_\nu$,
and obtain the form of the eigenstates
for the nonzero eigenvalues of $D_{\rm GW}^{m}$.
Since $(\Gamma^R+\hat{\Gamma})\sum_{\nu}c_{\nu}|J,J_3,\delta , \nu \rangle$
must be orthogonal to $\sum_{\nu}c_{\nu}|J,J_3,\delta , \nu \rangle$,
\beq
(a\alpha+2)|c_{1}|^2+(a\alpha-2)|c_{-1}|^2 =0.
\eeq
Therefore,  
the eigenstates for the eigenvalue $\alpha$ are
\beq
\frac{1}{2}\left[\sqrt{2-a\alpha}|J,J_3,\delta , 1 \rangle
 +\frac{\alpha}{|\alpha|}\sqrt{2+a\alpha}|J,J_3,\delta , -1 \rangle\right] ,
\eeq
where
$\alpha=
\pm \sqrt{\frac{n}{n+ m\delta}\left[ J(J+1)-\frac{m^2-1}{4}\right]}$.
Here we have absorbed the relative phase into 
the definition of $|J,J_3,\delta , 1 \rangle$
and $|J,J_3,\delta , -1 \rangle$.

Next, we check the index theorem (\ref{indextheorem})
by counting the number of the chiral zero-modes of $D_{\rm GW}^{m}$.
As noted before,  
the zero-modes for $D_{\rm GW}^{m}$ correspond to
the states with $J=\frac{m-1}{2}$,
whose degeneracy is $m$. 
Then, for $(\delta,\nu)=(+1,-1)$ in Table \ref{Mmeigens},
the index in the projected space is given by
\begin{equation}
{\rm{Index}}(P^{(n+ m)}D_{\rm GW})=n_{+}-n_{-}=0-m=-m.
\label{indexPnpmmm}
\end{equation}
For $(\delta,\nu)=(-1,+1)$, it is given by
\begin{equation}
{\rm{Index}}(P^{(n- m)}D_{\rm GW})=n_{+}-n_{-}=m-0=m.
\label{indexPnmmpm}
\end{equation}
Comparing with (\ref{TPpmm}),
this is consistent with the index theorem (\ref{indextheorem}).

Finally, we give some comments about the highest spin states.
As we noted above, 
the highest spin sates 
give only positive eigenvalue of $D_{\rm GW}^{m}$,
and thus the largest eigenvalue of the GW Dirac operator takes only 
a positive value (see the spectrum in Figure \ref{fig:spe}).
%\cite{Balachandran:2003ay}
This unbalance between the positive and the negative eigenvalues of $D_{\rm GW}$
comes from the fact that $D_{\rm GW}$ does not satisfy the ordinary
chiral symmetry, {\it i.e.} $\{ D_{\rm GW},\Gamma^R \} \neq 0$.
It is known that the highest spin states correspond to the species doublers
of the Watamura's Dirac operator $D_{\rm WW}$ \cite{Carow-Watamura:1996wg}.
%Then, in the present case, these states are
%would-be species doublers.
%\cite{balagovi,}

As can be seen from Table \ref{Mmeigens},
these modes have only $\nu = -1$,
which means that they have definite chirality 
defined by $\Gamma^R$.
The similar pattern of spectrum was provided for the GW Dirac operator $D$
in the lattice gauge theories \cite{Fujikawa:1999ku}:\\
i) $n_{\pm}$ states with eigenvalue zero for $\gamma_5 D$
and $\pm 1$ for the chirality $\gamma_5$.\\
ii) $N_{\pm}$ states with eigenvalue $\pm \frac{2}{a}$ for $\gamma_5 D$
and $\pm 1$ for the chirality $\gamma_5$.\\
iii) The remaining states have eigenvalues $\pm \alpha_n$ 
for $\gamma_5 D$
pairwise, with $0 < |\alpha_n| < \frac{2}{a}$.\\
It was also shown that the $N_{\pm}$ would-be species doubler sates 
play an important role in defining the index consistently.
As the index theorem states, in the well-defined continuum theories,
the trace of the chirality operator becomes
$\CTr_{\rm con}(\gamma_5)=n_{+} - n_{-}$,
while in the lattice theories
$\CTr(\gamma_5)=0$.
This discrepancy can be solved by 
taking account of the contributions from the $N_{\pm}$ 
would-be species doublers. 
Then $\CTr(\gamma_5)=n_{+} - n_{-} +N_{+}-N_{-} = 0$ is actually
satisfied on the lattice.
The would-be species doublers can be eliminated
by adopting 
$\frac{1}{2}\CTr(\gamma_5 +\hat{\gamma_5}) 
= \CTr(\gamma_5(1-\frac{1}{2}aD_{\rm GW}))$
instead of $\CTr(\gamma_5)$,
or by inserting some other factor which suppresses the contributions from
the large eigenvalues of the Dirac operator.
Then the correct value of the index is obtained,
and the smooth continuum limit can be taken.
Actually, these features of the spectrum and the index are given
only by the GW algebra, 
and they hold in the present case of the fuzzy 2-sphere as well.

Now let us come back to the case of the fuzzy 2-sphere,
and consider the counterpart of $\CTr(\gamma_5)$, which is
$\CTr(\Gamma^R)$.
From Table \ref{Mmeigens},
for the $\delta=1$ sector,
we can see that $n_{+} - n_{-} = -m$, $N_{+}-N_{-} = -(2n+m)$.
For the other states, both values of $\nu$ exist pairwise.
Thus
\beq
\CTr(P^{(n+m)}\Gamma^R)=(n_{+} - n_{-})+(N_{+}-N_{-})=-m-(2n+m)=-2(n+m).
\label{trpnpmgamr}
\eeq
In the same way, for the $\delta=-1$ sector, we see
\beq
\CTr(P^{(n-m)}\Gamma^R)=(n_{+} - n_{-})+(N_{+}-N_{-})=m-(2n-m)=-2(n-m).
\label{trpnmmgamr}
\eeq
We can eliminate the contributions from
the would-be species doublers 
%$(N_{+}-N_{-})$ 
by adopting 
$
\frac{1}{2}\CTr(P^{(n\pm m)}(\Gamma^R+\hat\Gamma))
=\CTr(P^{(n\pm m)}(\Gamma^R+\frac{1}{2}aD_{\rm GW}))
$,
as in the lattice gauge theories.
However,
(\ref{trpnpmgamr}) and (\ref{trpnmmgamr}) do not vanish,
while $\CTr(\gamma_5)=0$ in the lattice gauge theories.
This is because
the definition of $\Gamma^R$ in (\ref{gammaR})
has a constant term $-\frac{a}{2}=-\frac{1}{n}$, and thus
$\CTr(P^{(n\pm m)}\Gamma^R)$ has non-vanishing value, which is
$-\frac{1}{n}$ times the dimension of the Hilbert space,
$2n(n \pm m)$.
These non-vanishing values reflect the noncommutativity of the geometry
which can be interpreted as introducing a magnetic flux.

\section{Chiral zero-modes}\label{Chiralzero}
\setcounter{equation}{0}

In this section we will obtain the form of
the chiral zero-modes for the GW Dirac operator in the TP monopole background
with $m=1$.
By the unitary transformation (\ref{AiLtau}),
(\ref{gwdm}) becomes
\begin{eqnarray}
D_{\rm GW}^{\rm TP} &\equiv& U D_{\rm GW}^{m=1} U^\dagger \n
%           &=& U\!\begin{pmatrix}
%               \frac{n}{n+1}(\sigma_i L_i^{(n+1)} +\frac{1}{2}) & \cr
%               & \frac{n}{n-1}(\sigma_i L_i^{(n-1)} +\frac{1}{2}) \cr
%               \end{pmatrix}\!U^{\dagger} \!\!-(\sigma_i L_i^R -\frac{1}{2})\n
           &=& \sigma_i {\tilde L}_i +1 + \sigma_i\frac{\tau_i}{2}-
                 \frac{1}{n^2-1}L_i
                  \tau_i\Bigl[1+2\sigma_j(L_j+\frac{\tau_j}{2})\Bigr].
\label{gwdtp}
\end{eqnarray}
The benefit of taking this representation is 
that we can easily see the correspondence
between the noncommutative and commutative theories.
%For example, we can easily take the commutative limit.

As we showed in section \ref{spectDGW},
the zero-modes for $D_{\rm GW}^{\rm TP}$
correspond to the lowest spin states with $J=\frac{m-1}{2}$,
which is
$J=0$ for $m=1$.
%(See Table \ref{Mmeigens}.)
They thus can be written as
$\epsilon_{\alpha l}$ and $L^i\sigma^i_{\alpha \alpha'}\epsilon_{\alpha' l}$,
where $\alpha$ and $l$ are 
spinor and gauge group indices respectively.
Indeed, we can show directly
that these modes are zero-modes of $D_{\rm GW}^{\rm TP}$
by using the identity:
\begin{equation}
\sigma^i\ep=-\ep(\tau^{i})^{T},
\label{sigma2id1}
\end{equation}
or, if the indices written explicitly,
\beq
\sigma^i_{\alpha\alpha'}\ep_{\alpha' l}=-\tau^i_{ll'}\ep_{\alpha l'}.
\label{sigma2id2}
\eeq
For example,
\begin{eqnarray}
(D_{\rm GW}^{\rm TP}\ep)_{\alpha l}&=& \ep_{\alpha l}+\frac{1}{2}\sigma^i_{\alpha\alpha'}\tau^i_{ll'}\ep_{\alpha'l'}
                          -\frac{1}{n^2 -1}\bigl[ L^i\tau^i_{ll'}\ep_{\alpha l'}
                                           +2L^iL^j\sigma^j_{\alpha\alpha'}\tau^i_{ll'}\ep_{\alpha'l'}
                                         +L^i\sigma^j_{\alpha\alpha'}\tau^i_{ll'}\tau^j_{l'l''}\ep_{\alpha'l''}\bigr] \n
                     &=& \ep_{\alpha l}-\frac{1}{2}(\sigma^i\sigma^i\ep)_{\alpha l}
                         -\frac{1}{n^2 -1}\bigl[ -L^i(\sigma^i\ep)_{\alpha l} 
                                          -2L^iL^j(\sigma^j\sigma^i\ep)_{\alpha l}
                                          +L^i(\sigma^j\sigma^j\sigma^i\ep)_{\alpha l}\bigr] \n
                     &=& \ep_{\alpha l}-\frac{3}{2}\ep_{\alpha l}
                         -\frac{1}{n^2 -1}\Bigl[ 2L^i(\sigma^i\ep)_{\alpha l} 
                                          -\frac{n^2-1}{2}\ep_{\alpha l}
                                          -2L^i(\sigma^i\ep)_{\alpha l}\Bigr] \n
                     &=& 0 .
\end{eqnarray}
In the same way,
we can show 
$(D_{\rm GW}^{\rm TP}L^i\sigma^i\ep)_{\alpha l}= 0$. 

%The simultaneous eigenstates of $T$ and $\Gamma^R$
%in the zero-mode space 
Chiral zero-modes can be obtained by
the linear combinations of these two zero-modes
$\epsilon_{\alpha l}$ and $L^i\sigma^i_{\alpha \alpha'}\epsilon_{\alpha' l}$.
The states with $(\delta,\nu)=(\mp 1,\pm 1)$ are given by
\beq
|J=0, \delta=\mp 1, \nu=\pm 1> = 
\frac{1}{2}(1 \pm \Gamma^R) \ep
=\frac{1}{2}(1 \mp T) \ep,
\eeq
or, if the indices written explicitly,
\beq
|J=0, \delta=\mp 1, \nu=\pm 1> =
\frac{1}{2}\Bigl[(1 \mp \frac{1}{n}) \ep_{\alpha l} 
\pm \frac{2}{n} L^i \sigma^i_{\alpha \alpha'} \ep_{\alpha' l}\Bigr]
=\frac{1}{2}\Bigl[(1 \mp \frac{1}{n}) \ep_{\alpha l} 
\mp \frac{2}{n} L^i \tau^i_{l l'} \ep_{\alpha l'}\Bigr],
\label{chizeromononcom}
\eeq
where again we used the identity (\ref{sigma2id2}).

In appendix \ref{sec:speccom}, we calculate the spectrum of the 
Dirac operator in the commutative theory.
We also obtain the chiral zero-modes in
(\ref{chiralzeromodescom}): 
\begin{equation}
|J=0, \delta=\mp 1, \nu=\pm 1>_{\rm com} =
\frac{1}{2}[1\pm n^i \sigma^i]_{\alpha\alpha'}\ep_{\alpha' l}
=\frac{1}{2}[1\mp n^i \tau^i]_{ll'}\ep_{\alpha l'},
\label{chizeromocom}
\end{equation}
where $n^i = x^i /\rho$ is a unit vector.
As we mentioned before,
the correspondence between
the noncommutative and commutative theories
can be easily seen
in the representation (\ref{gwdtp}).
Indeed, we can see that the commutative limit of (\ref{chizeromononcom})
becomes (\ref{chizeromocom}).

\section{Conclusions and Discussions}
\setcounter{equation}{0}

In this paper,
we showed the index theorem for the Ginsparg-Wilson Dirac operator 
in the 't Hooft Polyakov monopole backgrounds
and provided the meaning of the projection operator.
We then calculated the spectrum and eigenstates, 
%for the Dirac operator in these backgrounds,
and confirmed the index theorem
by counting the number of the chiral zero-modes.
We also showed that
the largest-eigenvalue modes
%of the GW Dirac operator has only 
%a positive value,
%and that these modes have definite chirality. 
%These would-be species doublers 
play an important role in defining the index consistently
in the theories with finite degrees of freedom.

%We have also
%provided a physical meaning of the projection operator $P^{(n\pm m)}$
%which is necessary to formulate
%the index theorem in the TP monopole backgrounds.
%Namely, the TP monopole breaks $SU(2)$ gauge symmetry
%down to $U(1)$,
%and the matter field in the fundamental representation has 
%two components, corresponding to $\pm 1/2$ electric charge 
%components of the unbroken $U(1)$ gauge group.
%The projector $P^{(n\pm m)}$ picks up one of the two components.
%With the projection operator, 
%nonzero index is obtained for the TP monopole backgrounds.
%Without the projector,
%the contributions from the two components cancel the index. 

One of the mail results of the paper is that
we have extended the index theorem to general configurations
which are not restricted to the special type of solutions.
By this generalization, configuration space can be classified 
into topological sectors.
The commutative limit of the topological charge 
becomes the one introduced by 't Hooft in spontaneously
 broken gauge theories.
We also considered the condition to assure the validity 
of this formulation,
which gives the generalization of the admissibility conditions
in the lattice gauge theory.
Since this formulation is gauge invariant,
it might be used to formulate
the chiral gauge theory.
% on the fuzzy 2-sphere.
Abelian chiral gauge theories on the lattice with exact gauge invariance
was constructed
by using the chiral projection operator \cite{Luscher:1998kn}.
It is an interesting future problem
whether a generalization of our formulation
provides an alternative to it.

It is also interesting to study
the TP monopole configurations of general $m$
in the commutative and noncommutative theories\footnote
{Some comments are given in \cite{ISTT}.
}.
%, they study the TP monopole configurations
%with general $m$,
%and investigate the interaction of the fluctuation around the 
%monopole backgrounds by harmonic expansion.
%%We will study the forms of configurations themselves.
%}.
By some gauge transformations,
TP monopole configurations can be seen as 
Dirac monopoles 
or Wu-Yang monopoles where we need to introduce the notion of patch.
Hence this study will lead to a formulation of monopole bundles in the 
noncommutative geometries or matrix models.
Also it is interesting to study whether configurations with
a nontrivial index exist without introducing
the projection operator.
In a discretized noncommutative torus,
nontrivial configurations can exist without introducing a projection operator,
though their existence probability vanishes in the continuum limit \cite{Nagao:2005st, ANS}.
These studies may provide another meaning of the projection operator.
%We will report on these studies in a separate paper in future. 

\section*{Acknowledgements}
We would like to thank K. Funakubo, H. Kawai,
J. Nishimura and M. Sakamoto, A. Tsuchiya
for discussions and useful comments.

\appendix

\section{Spectrum of the GKP Dirac operator}
\label{sec:specGKP}
\setcounter{equation}{0}

In this appendix, we will obtain the spectrum for
$D_{\rm GKP}$ (\ref{DGKP}) in the TP monopole backgrounds
({\ref{LnpmLnmm}):
\begin{equation}
D_{\rm GKP}^m =\sigma_i(A_i-L_i^R) +1 
              =\sigma_i\left[ \begin{pmatrix}
                 L_i^{(n+m)} & \cr
                 & L_i^{(n-m)}  \cr
                 \end{pmatrix}-L_i^R \right] +1
\end{equation}
Note $L_i^{(n \pm m)}$ has spin $L \pm m/2$, 
and $-L_i^R$ has spin $L$, where $n= 2L+1$.
Then the operator $A_i-L_i^R$
has the following spins:
\begin{equation}
l=\left\{
\begin{array}{cc}
\frac{m}{2},\cdots,2L+\frac{m}{2}& (\delta = 1),\\
\frac{m}{2},\cdots,2L-\frac{m}{2}& (\delta = -1).
\end{array}\right.
\label{laili}
\end{equation}
Here we set $m \ge 0$.
Then, by  considering the spin $J$ 
for the operator $M_i$ of (\ref{defMA}),
we obtain the following spectrum for $D_{\rm GKP}^{m}$:
\begin{equation}
D_{\rm GKP}^{m}
= \left\{
\begin{array}{ccc}
J+\frac{1}{2}& =&l+1\ \ \ (J=l+\frac{1}{2}), \\
-(J+\frac{1}{2})& =&-l\ \ \ (J=l-\frac{1}{2}),   
\end{array}\right.
\end{equation}
where $l$'s are given by (\ref{laili}).
The spectrum is shown in Figure \ref{GKPmspe} for $m>0$ cases.
For $m=0$, the spectrum is given by Figure \ref{GKPmspe}
except for the fact that zero-modes do not exist.
We also illustrate the cases for $n=10, m=0,1,2,3,4$ 
in Figure \ref{fig:GKP012spe}.
\begin{figure}[h]
\begin{center}
\begin{picture}(150,300)(,)
\put(10,290){$D_{\rm GKP}$}
\put(20,-20){\vector(0,1){295}}
\put(20,250){\line(1,0){80}}
%\put(20,230){\line(1,0){80}}
\put(20,210){\line(1,0){80}}
\put(20,170){\line(1,0){80}}
\put(20,150){\line(1,0){80}}
%\put(20,130){\line(1,0){80}}
\put(20,115){\line(1,0){90}}
\put(20,100){\line(1,0){80}}
\put(20,80){\line(1,0){80}}
\put(20,50){\line(1,0){80}}
%\put(20,30){\line(1,0){80}}
\put(20,10){\line(1,0){80}}

\put(-44,247){$2L+\frac{m}{2}+1$}
\put(-2,224){\rotatebox{90}{$\cdots$}}
\put(84,224){\rotatebox{90}{$\cdots$}}
\put(-44,207){$2L-\frac{m}{2}+1$}
\put(-2,186){\rotatebox{90}{$\cdots$}}
\multiput(39,186)(45,0){2}{\rotatebox{90}{$\cdots$}}
\put(-18,167){$\frac{m}{2}+2$}
\put(-18,147){$\frac{m}{2}+1$}
\put(-6,112){$0$}
\put(-16,97){$-\frac{m}{2}$}
\put(-35,77){$-(\frac{m}{2}+1)$}
%\put(-15,60){$-5/2$}
\put(-8,59){\rotatebox{90}{$\cdots$}}
\multiput(40,59)(45,0){2}{\rotatebox{90}{$\cdots$}}
\put(-41,46){$-(2L-\frac{m}{2})$}
\put(-8,25){\rotatebox{90}{$\cdots$}}
\put(85,25){\rotatebox{90}{$\cdots$}}
\put(-41,6){$-(2L+\frac{m}{2})$}
\put(87,250){\circle{5}}
%\put(57,230){\circle{5}}
\multiput(42,210)(45,0){2}{\circle{5}}
\multiput(42,170)(45,0){2}{\circle{5}}
\multiput(42,150)(45,0){2}{\circle{5}}
\multiput(42,100)(45,0){2}{\circle{5}}
\multiput(42,80)(45,0){2}{\circle{5}}
\multiput(42,50)(45,0){2}{\circle{5}}
%\multiput(32,30)(45,0){2}{\circle{5}}
\put(87,10){\circle{5}}
\put(30,270){$\delta=-1$}
\put(80,270){$\delta=1$}
\end{picture}
\end{center}
\caption{Spectrum for the GKP Dirac operator in the monopole backgrounds
of $m> 0$.}
\label{GKPmspe}
\end{figure}
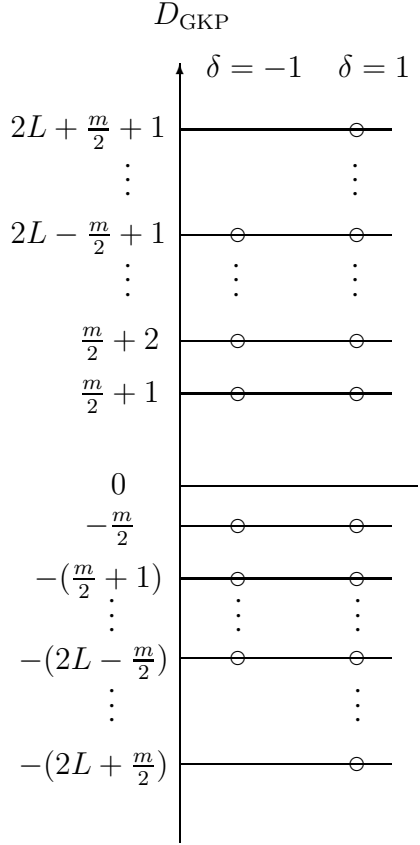
\begin{figure}[hp]
\begin{center}
\includegraphics[height=20cm]{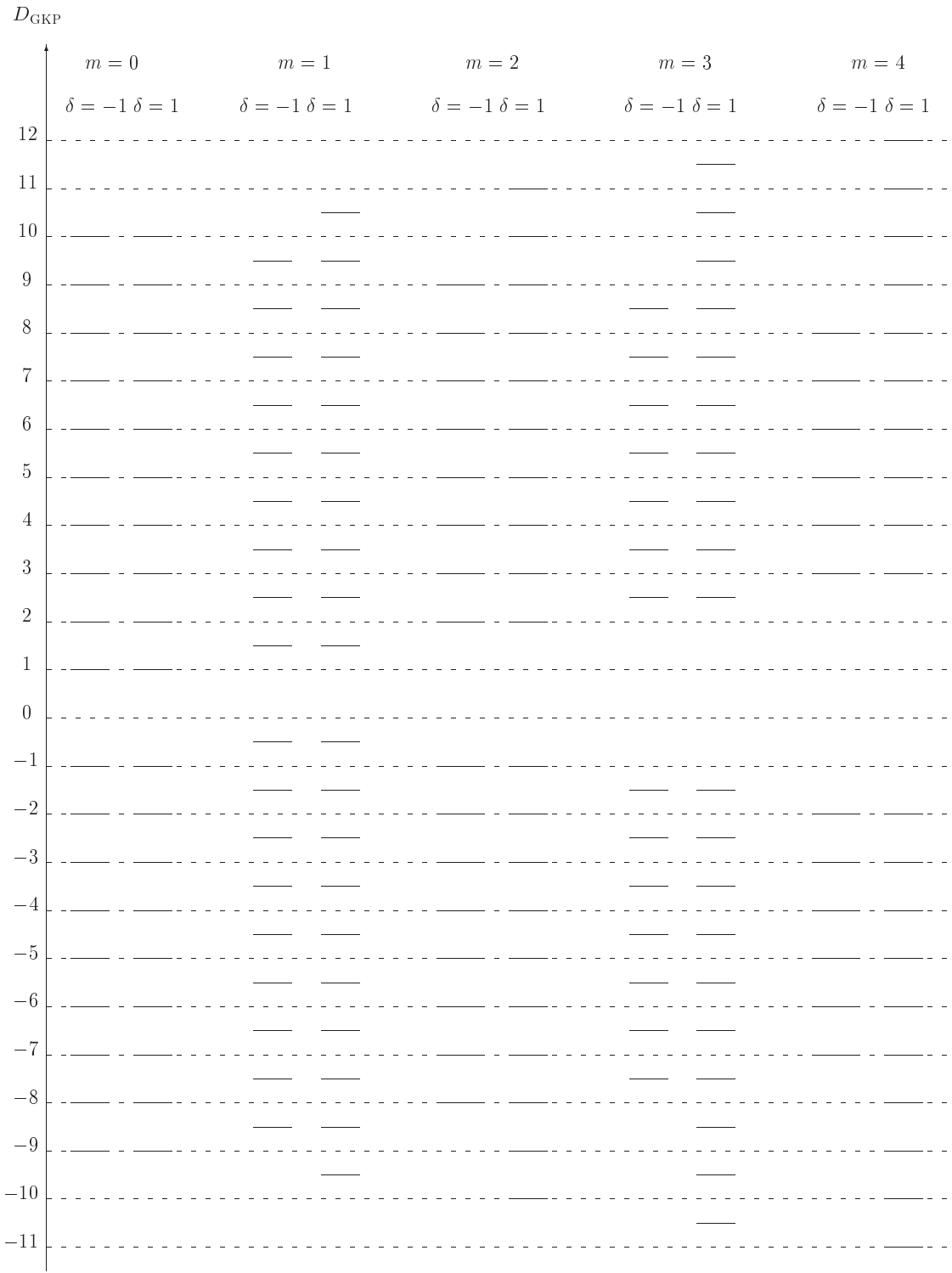}
\end{center}
\caption{Spectrum for the GKP Dirac operator for the cases of $n=10$ and
$m=0,1,2,3,4$.}
\label{fig:GKP012spe}
\end{figure}

In the remainder of this appendix, 
we will obtain the above spectrum in another way,
as we did for $D_{\rm GW}^m$ in section \ref{spectDGW}.
Then we can make the correspondence of the eigenstates for $D_{\rm GKP}^m$
to the eigenstates in Table \ref{Mmeigens}.
In particular, we can see the chirality of $\Gamma^R$, $\nu$, of
the $D_{\rm GKP}^m$ eigenstates.
By straightforward calculations, we can show
\begin{equation}
(D_{\rm GKP}^m)^2 = M_i^2+\frac{1}{4}=J(J+1)+\frac{1}{4}=(J+\frac{1}{2})^2,
\end{equation}
and $(D_{\rm GKP}^m)^2$ commutes with $M_i, T, \Gamma^R$.
%and then
%\beqa
%\left[ M_i \  ,\  (D_{\rm GKP}^{m})^2\right] &=& 0, \\
%\left[ T \  ,\  (D_{\rm GKP}^{m})^2\right] &=& 0, \\
%\left[ \Gamma^R \  ,\  (D_{\rm GKP}^{m})^2\right] &=& 0.  
%\eeqa
Thus  
spectrum for $(D_{\rm GKP}^m)^2$ is
\beq
(D_{\rm GKP}^{m})^2|J,J_3,\delta , \nu \rangle =
(J+\frac{1}{2})^2
 \ |J,J_3,\delta , \nu \rangle,
\eeq
where the simultaneous eigenstate
$|J,J_3,\delta , \nu \rangle$ is 
defined in (\ref{sesmm})-(\ref{sesgammar}).

Since $D_{\rm GKP}^{m}$ satisfies
\beqa
\left[ M_i \  ,\  D_{\rm GKP}^{m}\right] &=& 0, \\
\left[ T \  ,\  D_{\rm GKP}^{m}\right] &=& 0, \\
\left[ \Gamma^R \  ,\  D_{\rm GKP}^{m}\right] &\ne& 0,
\eeqa
linear combinations over $\nu$ for each fixed $J,J_3,\delta$,
$\sum_\nu c_\nu |J,J_3,\delta , \nu \rangle$,
give eigenstates for $D_{\rm GKP}^{m}$ as
\beq
D_{\rm GKP}^{m}\sum_\nu c_\nu |J,J_3,\delta , \nu \rangle =
\pm(J+\frac{1}{2})
 \ \sum_\nu c_\nu |J,J_3,\delta , \nu \rangle.
 \label{esDGKPlc}
\eeq

We now define
\begin{equation}
D_{\rm GKP}^{m \pm} = 
(\sigma_i L_i^{(n\pm m)}+\frac{1}{2})-(\sigma_i L_i^R-\frac{1}{2}) 
= \frac{n\pm m}{2}\hat\Gamma^\pm -\frac{n}{2}\Gamma^R,
\label{dgkpmpm}
\end{equation}
where
\begin{eqnarray}
\Gamma^R&=&\frac{2}{n}(\sigma_iL_i^R-\frac{1}{2}), \\
\hat\Gamma^\pm&=&\frac{2}{n\pm m}(\sigma_iL_i^{(n\pm m)}+\frac{1}{2}),
\end{eqnarray}
in each of $\delta= \pm 1$ sectors.
Then we can show the following Ginsparg-Wilson-like relation:
\beqa
\frac{n\pm m}{n}\hat\Gamma^\pm {D_{\rm GKP}^{m\pm}} 
+D_{\rm GKP}^{m\pm}\Gamma^R&=&\pm m + \frac{m^2}{2n}, 
\label{GKPgwrel1}\\
\frac{n\pm m}{n}D_{\rm GKP}^{m\pm}\hat\Gamma^\pm+\Gamma^R D_{\rm GKP}^{m\pm}
&=&\pm m + \frac{m^2}{2n}.
\label{GKPgwrel2} 
\eeqa
Suppose $\psi$ is an
eigenstate for $D_{\rm GKP}^{m\pm}$ with an eigenvalue $\alpha$:
\begin{equation}
D_{\rm GKP}^{m\pm}\psi =\alpha \psi.
\end{equation}
Then, from (\ref{GKPgwrel1}) and (\ref{GKPgwrel2}),
\begin{equation}
D_{\rm GKP}^{m\pm}\left(\Gamma^R+\frac{n\pm m}{n}\hat\Gamma^\pm\right)\psi 
=-\alpha \left(\Gamma^R+\frac{n\pm m}{n}\hat\Gamma^\pm\right)\psi
+ \left(\pm 2m + \frac{m^2}{n}\right)\psi.
\end{equation}
Thus,
\beqa
&&D_{\rm GKP}^{m\pm}
\left[\Gamma^R+\frac{n\pm m}{n}\hat\Gamma^\pm
-\frac{1}{2\alpha}\left(\pm 2m + \frac{m^2}{n}\right)\right]\psi \nonumber \\
&=&-\alpha
\left[\Gamma^R+\frac{n\pm m}{n}\hat\Gamma^\pm-
\frac{1}{2\alpha}\left(\pm 2m + \frac{m^2}{n}\right)\right]\psi.
\eeqa
Hence,
$\left[\Gamma^R+\frac{n\pm m}{n}\hat\Gamma^\pm
-\frac{1}{2\alpha}\left(\pm 2m + \frac{m^2}{n}\right)\right]\psi$
is an eigenstate for $D_{\rm GKP}^{m\pm}$
with an eigenvalue $-\alpha$.

From (\ref{dgkpmpm}),
\beq
\Gamma^R+\frac{n\pm m}{n}\hat\Gamma^\pm
=\frac{2}{n}{D_{\rm GKP}^{m\pm}}+2\Gamma^R.
\eeq
Then, for the eigenstate 
$\psi=\sum_\nu c_\nu |J,J_3,\delta , \nu \rangle$
in (\ref{esDGKPlc}}),
\begin{eqnarray}
&&\left[\Gamma^R+\frac{n +\delta m}{n}\hat\Gamma
-\frac{1}{2\alpha}\left(2\delta m + \frac{m^2}{n}\right)\right]
\sum_\nu c_\nu |J,J_3,\delta , \nu \rangle \nonumber\\
&=&\left[\frac{2}{n}\alpha-2
-\frac{1}{2\alpha}\left( 2\delta m+ \frac{m^2}{n}\right)\right]
c_{-1}|J,J_3,\delta , \nu=-1 \rangle \n
&&+\left[\frac{2}{n}\alpha+2
-\frac{1}{2\alpha}\left( 2\delta m+ \frac{m^2}{n}\right)\right]
c_1|J,J_3,\delta , \nu=1 \rangle,
\label{meves}
\end{eqnarray}
where $\alpha= \pm (J+ \frac{1}{2})$.
For the lowest spin states with $J=\frac{m-1}{2}$ in Table \ref{Mmeigens},
since only $\nu=+1$ or $\nu=-1$ exists for each $\delta$,
the state (\ref{meves}) must vanish.
Thus we can show $\alpha = -\frac{m}{2}$.
Hence only negative eigenvalue for $D_{\rm GKP}^{m}$
exist for the lowest spin states.
For the highest spin states,
$(J,\delta,\nu)=(2L-\frac{m-1}{2}, -1,-1)$,
$(2L+\frac{m+1}{2}, +1,-1)$, 
only $\nu=-1$ exists for each $\delta$,
and the state (\ref{meves}) must vanish.
Thus we can show $\alpha = n \pm \frac{m}{2}$ for $\delta = \pm 1$.
Hence only positive eigenvalues for $D_{\rm GKP}^{m}$
exist for the highest spin states.
For the other states, 
%(\ref{meves}) do not vanish, and thus 
both positive and  negative eigenvalues exit.
Then we obtain the spectrum in figure \ref{GKPmspe} again.

We  will also determine the coefficients $c_\nu$, 
and obtain the form of the eigenstates.
Since the state (\ref{meves}) and 
$\sum_\nu c_\nu |J,J_3,\delta , \nu \rangle$ are orthogonal,
\beq
\left[\frac{2}{n}\alpha-2
-\frac{1}{2\alpha}\left( 2\delta m+ \frac{m^2}{n}\right)\right]
|c_{-1}|^2
+\left[\frac{2}{n}\alpha+2
-\frac{1}{2\alpha}\left( 2\delta m+ \frac{m^2}{n}\right)\right]
|c_1|^2=0
\eeq
Therefore the eigenstates for $D_{\rm GKP}^{m}$ is given by
\beqa
&&\frac{1}{2}\Biggl[\sqrt{2+\frac{2}{n}\alpha
-\frac{1}{2\alpha}\left( 2\delta m+ \frac{m^2}{n}\right)}
|J,J_3,\delta , \nu=-1 \rangle \n
&&+\frac{\alpha}{|\alpha|}
\sqrt{2-\frac{2}{n}\alpha
+\frac{1}{2\alpha}\left( 2\delta m+ \frac{m^2}{n}\right)}
|J,J_3,\delta , \nu=1 \rangle \Biggr],
\eeqa
where $\alpha= \pm(J+\frac{1}{2})$. 
Here we have absorbed the relative phase into the
definition of  
$|J,J_3,\delta , \nu=-1 \rangle$ and
$|J,J_3,\delta , \nu=1 \rangle$.

\section{Spectrum of the Dirac operator in the commutative theory}
\label{sec:speccom}
\setcounter{equation}{0}

In this appendix,
we will calculate the spectrum for the Ginsparg-Wilson Dirac operator
in the commutative theory (\ref{DGWcom})
in the TP monopole background (\ref{monopoleconfig}) with $m=1$:
\begin{equation}
D_{\rm com}^{\prime \ m=1} = \sigma\cdot {\cal L} +1 + \frac{1}{2}\sigma\cdot\tau-\frac{1}{2}(n\cdot\sigma)(n\cdot\tau).
\end{equation}
It is also obtained by the commutative limit of (\ref{gwdtp}).

We consider the following operators:
\begin{eqnarray}
M_i&=&{\cal L}_i + \frac{\sigma_i}{2} +\frac{\tau_i}{2}, \label{totall}\\
t&=&n\cdot\tau, \\
\gamma &=& n\cdot\sigma,
\end{eqnarray}
where $M_i$ is a total angular momentum operator,
$t$ is a generator for the unbroken $U(1)$ gauge group,
and $\gamma$ is a chirality operator. 
Since they commute with one another,
%\begin{eqnarray}
%\left[ M_i \ ,\  t \right] &=& 0, \\
%\left[ M_i \  ,\  \gamma \right] &=& 0, \\
%\left[ t \  ,\  \gamma \right] &=& 0, 
%\end{eqnarray}
we can consider the simultaneous eigenstates for these operators as
\begin{eqnarray}
M_i^2|J,J_3,\delta , \nu \rangle &=&J(J+1) \ |J,J_3,\delta , \nu \rangle, \\
M_3|J,J_3,\delta , \nu \rangle &=&J_3 \ |J,J_3,\delta , \nu \rangle, \\
t|J,J_3,\delta , \nu \rangle &=&\delta \ |J,J_3,\delta , \nu \rangle, \\
\gamma|J,J_3,\delta , \nu \rangle &=&\nu \ |J,J_3,\delta , \nu \rangle. 
\end{eqnarray}
%
%By considering a operator ${\cal L}_i + \frac{\sigma_i}{2}$, and 
%then 
%$M_i = ({\cal L}_i + \frac{\sigma_i}{2}) +\frac{\tau_i}{2}$,
%we obtain the following spins:
%\begin{equation}
%J=\left\{
%\begin{array}{cc}
%l+1& (l \geq 0),\\
%l& (l \geq 0),\\
%l& (l \geq 1),\\
%l-1& (l \geq 1),
%\end{array}\right.
%\label{Mi}
%\end{equation}
%where $l$ is the spin for the operator ${\cal L}_i$.
$J$ turns out to take values as in the Table \ref{comDeigs}
\footnote{
As we will see later in (\ref{chiralzeromodescom}), 
two zero-modes exist
as $(\delta,\nu)=(-1,+1)$ and $(+1,-1)$.
From (\ref{MiD}), (\ref{tD}), (\ref{antigD}), 
the eigenstates  with both $\nu =+1$ and $\nu =-1$ 
exist in each $J,J_3,\delta$
for $J\geq 1$.
(Also, by defining 
$D' = \tau\cdot {\cal L} +1 + \frac{1}{2}\sigma\cdot\tau-\frac{1}{2}(n\cdot\sigma)(n\cdot\tau)$, 
we can show that the eigenstates with both $\delta =+1$ and $\delta =-1$
exist in each $J,J_3,\nu$
for $J\geq 1$.)
We can also show that there is no degeneracy in each
$|J,J_3,\delta , \nu \rangle$,
by counting the number of states in each $J$.
}.
\begin{table}[t]
\caption{Values of $J$ in each $\delta$, $\nu$ sector.}
\label{comDeigs}
\begin{center}
\begin{tabular}{|c@{\quad\vrule width0.8pt\quad}ccccc|} 
\hline
         &  $\delta $     & $-$     & $-$     & $+$     & $+$     \\
$J$      &  $\nu $ & $+$     & $-$     & $+$     & $-$     \\
\noalign{\hrule height 0.8pt}
$0$      &                    & $\circ$ &         &         & $\circ$ \\
$1$      &                    & $\circ$ & $\circ$ & $\circ$ & $\circ$ \\
$2$      &                    & $\circ$ & $\circ$ & $\circ$ & $\circ$ \\
$3$      &                    & $\circ$ & $\circ$ & $\circ$ & $\circ$ \\
\rotatebox{90}{$\cdots$} &                
                              &\rotatebox{90}{$\cdots$} 
                                        &\rotatebox{90}{$\cdots$} 
                                                  &\rotatebox{90}{$\cdots$} 
                                                            &\rotatebox{90}{$\cdots$} \\
\hline
\end{tabular}
\end{center}
\end{table}

By straightforward calculations,
square of $D_{\rm com}^{\prime \ m=1}$ becomes
\begin{equation}
(D_{\rm com}^{\prime  \ m=1})^2= M_i^2,
\end{equation}
and $(D_{\rm com}^{\prime  \ m=1})^2$
commutes with $M_i, t, \gamma$.
%\begin{eqnarray}
%\left[ M_i \  ,\  (D_{\rm com}^{\prime  \ m=1})^2 \right] &=& 0, \\
%\left[ t \  ,\  (D_{\rm com}^{\prime  \ m=1})^2 \right] &=& 0, \\
%\left[ \gamma \  ,\  (D_{\rm com}^{\prime  \ m=1})^2 \right] &=& 0. 
%\end{eqnarray}
We thus obtain the spectrum for $(D_{\rm com}^{\prime  \ m=1})^2$ as follows:
\beq
(D_{\rm com}^{\prime  \ m=1})^2 \ |J,J_3,\delta , \nu \rangle =
 J(J+1) \ |J,J_3,\delta , \nu \rangle.
\eeq
Note that the states with the lowest spin, $J=0$,
in figure \ref{comDeigs}
correspond to the zero-modes for
$D_{\rm com}^{\prime  \ m=1}$.

We can also show
\begin{eqnarray}
\left[ M_i \  ,\  D_{\rm com}^{\prime  \ m=1} \right] &=& 0, \label{MiD}\\
\left[ t \  ,\  D_{\rm com}^{\prime  \ m=1} \right] &=& 0, \label{tD}\\
\left[ \gamma \  ,\  D_{\rm com}^{\prime  \ m=1} \right] &\ne & 0. \label{gD}
\end{eqnarray}
Then, linear combinations over $\nu$ for each $J,J_3,\delta $,
$\sum_{\nu}c_{\nu}|J,J_3,\delta , \nu \rangle$, 
give eigenstates for the Dirac operator $D_{\rm com}^{\prime  \ m=1}$ as
\begin{equation}
D_{\rm com}^{\prime  \ m=1}
\left[\sum_{\nu}c_{\nu}|J,J_3,\delta , \nu \rangle\right] 
=\pm \sqrt{J(J+1)}
\left[\sum_{\nu}c_{\nu}|J,J_3,\delta , \nu \rangle\right] . 
\label{esgwd}  
\end{equation}

Since Dirac operator $D_{\rm com}^{\prime  \ m=1}$ anti-commutes with the chirality operator $\gamma$:
\begin{eqnarray}
\left\{ \gamma \  ,\  D_{\rm com}^{\prime  \ m=1} \right\} &=& 0 ,
\label{antigD}
\end{eqnarray}
if $\sum_{\nu}c_{\nu}|J,J_3,\delta , \nu \rangle$ 
is an eigenstate for the $D_{\rm com}^{\prime  \ m=1}$ with eigenvalue $\alpha$,
$\gamma\sum_{\nu}c_{\nu}|J,J_3,\delta , \nu \rangle$
is an eigenstate with eigenvalue $-\alpha$.
%\beqa
%D_{\rm com}^{\prime  \ m=1}
%\left[\sum_{\nu}c_{\nu}|J,J_3,\delta , \nu \rangle\right] 
%&=&
%\alpha\left[\sum_{\nu}c_{\nu}|J,J_3,\delta , \nu \rangle\right],\\
%D_{\rm com}^{\prime  \ m=1}
%\left[\gamma\sum_{\nu}c_{\nu}|J,J_3,\delta , \nu \rangle\right] 
%&=&
%-\alpha\left[\gamma\sum_{\nu}c_{\nu}|J,J_3,\delta , \nu \rangle\right],
%\eeqa
%where
%\beq
%\alpha=
%\pm \sqrt{J(J+1)}.
%\label{dgweigenvalue}
%\eeq
Since
\beq
\gamma\sum_{\nu}c_{\nu}|J,J_3,\delta , \nu \rangle=
-c_{-1}|J,J_3,\delta , \nu=-1 \rangle
+c_{1}|J,J_3,\delta , \nu=1 \rangle
\label{gammalccom}
\eeq
does not vanish,
both positive and negative eigenvalues for $D_{\rm com}^{\prime  \ m=1}$
exist in each $J \ge 1$ and $\delta$.

We next determine the coefficients $c_\nu$,
and obtain the form of the eigenstates.
Since the state (\ref{gammalccom})
is orthogonal to $\sum_{\nu}c_{\nu}|J,J_3,\delta , \nu \rangle$,
\beq
|c_{1}|^2-|c_{-1}|^2 =0.
\eeq
Therefore,  
the eigenstates for the eigenvalue $\alpha$ are
\beq
\frac{1}{\sqrt{2}}\left[ |J,J_3,\delta , 1 \rangle
 +\frac{\alpha}{|\alpha|}|J,J_3,\delta , -1 \rangle\right],
\eeq
where $\alpha=\pm \sqrt{J(J+1)}$.
Here we have absorbed the relative phase into 
the definition of $|J,J_3,\delta , 1 \rangle$
and $|J,J_3,\delta , -1 \rangle$.

Finally, we obtain configuration form of 
the chiral zero-modes.
Since they correspond to the zero-modes of 
$M_i$, the modes with $J=0$ in Table \ref{comDeigs}, 
they can be written as
$\epsilon_{\alpha l}$ and $n^i\sigma^i_{\alpha \alpha'}\epsilon_{\alpha' l}$,
where $\alpha$ and $l$ are 
spinor and gauge group indices respectively.
We can indeed show $(D_{\rm com}^{\prime  \ m=1}) \ep=0$,
$(D_{\rm com}^{\prime  \ m=1}) n \cdot \sigma \ep =0$ directly. 
Chiral zero-modes can be obtained by
the linear combinations of these two zero-modes.
The states with $(\delta,\nu)=\pm(-1,+1)$ are
\begin{equation}
\frac{1}{2}[1\pm n\cdot\sigma]_{\alpha\alpha'}\ep_{\alpha' l}=\frac{1}{2}[1\mp n\cdot\tau]_{ll'}\ep_{\alpha l'} ,
\label{chiralzeromodescom}
\end{equation}
where we used the identity (\ref{sigma2id2}).
 
The eigenvalues and eigenstates obtained here
agree with
the commutative limit of 
the results in section  \ref{spectDGW} and section \ref{Chiralzero}.
We also note that the monopole harmonics in the commutative theory 
was provided in \cite{Wu:1976ge},
and the spectrum of the equivalent Dirac operator 
was studied in
\cite{Rubakov:1982fp}.


\begin{thebibliography}{99}
\bibitem{Banks:1996vh}
T.~Banks, W.~Fischler, S.~H.~Shenker and L.~Susskind,
%``M theory as a matrix model: A conjecture,''
Phys.\ Rev.\ D {\bf 55}, 5112 (1997)
[arXiv:hep-th/9610043].
%%CITATION = HEP-TH 9610043;%%
 
\bibitem{IKKT}
N.~Ishibashi, H.~Kawai, Y.~Kitazawa and A.~Tsuchiya,
%``A large-N reduced model as superstring,''
Nucl.\ Phys.\ B {\bf 498}, 467 (1997)
[arXiv:hep-th/9612115].
%%CITATION = HEP-TH 9612115;%%

For a review, see
H.~Aoki, S.~Iso, H.~Kawai, Y.~Kitazawa, A.~Tsuchiya and T.~Tada,
%``IIB matrix model,''
Prog.\ Theor.\ Phys.\ Suppl.\  {\bf 134}, 47 (1999)
[arXiv:hep-th/9908038].
%%CITATION = HEP-TH 9908038;%%
 
\bibitem{Connes}
A. Connes, Noncommutative geometry, Academic Press, 1990.
 
\bibitem{CDS} A.~Connes, M.~R.~Douglas and A.~Schwarz,
%``Noncommutative geometry and matrix theory: Compactification on tori,''
JHEP {\bf 9802}, 003 (1998)
[arXiv:hep-th/9711162].
%%CITATION = HEP-TH 9711162;%%
 
\bibitem{NCMM}
H.~Aoki, N.~Ishibashi, S.~Iso, H.~Kawai, Y.~Kitazawa and T.~Tada,
%``Noncommutative Yang-Mills in IIB matrix model,''
Nucl.\ Phys.\ B {\bf 565}, 176 (2000)
[arXiv:hep-th/9908141].
%%CITATION = HEP-TH 9908141;%%
 
\bibitem{Seiberg:1999vs}
 N.~Seiberg and E.~Witten,
 %``String theory and noncommutative geometry,''
 JHEP {\bf 9909} (1999) 032
  [arXiv:hep-th/9908142].
  %%CITATION = HEP-TH 9908142;%%


%-----------------------Non-trivial config.-----------------
\bibitem{non-trivial_config}
%\bibitem{23_5}
H.~Grosse, C.~Klimcik and P.~Presnajder,
%``Topologically nontrivial field configurations in noncommutative
%geometry,''
Commun.\ Math.\ Phys.\  {\bf 178}, 507 (1996)
[arXiv:hep-th/9510083];
%%CITATION = HEP-TH 9510083;%%
 
S.~Baez, A.~P.~Balachandran, B.~Ydri and S.~Vaidya,
%``Monopoles and solitons in fuzzy physics,''
Commun.\ Math.\ Phys.\  {\bf 208}, 787 (2000)
[arXiv:hep-th/9811169];
 
G.~Landi,
%``Projective modules of finite type and monopoles over S(2),''
J.\ Geom.\ Phys.\  {\bf 37}, 47 (2001)
[arXiv:math-ph/9905014].
%%CITATION = MATH-PH 9905014;%%
 
%-----------non-trivial config.and chiral anomaly-------
\bibitem{non_chi}
A.~P.~Balachandran and S.~Vaidya,
%``Instantons and chiral anomaly in fuzzy physics,''
Int.\ J.\ Mod.\ Phys.\ A {\bf 16}, 17 (2001)
[arXiv:hep-th/9910129].
%%CITATION = HEP-TH 9910129;%%
%------------------------------------------------------------
 
%--------------------------------------------------------------
%\cite{Valtancoli:2001gx}
\bibitem{Valtancoli:2001gx}
P.~Valtancoli,
%``Projectors for the fuzzy sphere,''
Mod.\ Phys.\ Lett.\ A {\bf 16}, 639 (2001)
[arXiv:hep-th/0101189].
%%CITATION = HEP-TH 0101189;%%
 
 
%\cite{Steinacker:2003sd}
\bibitem{Steinacker:2003sd}
  H.~Steinacker,
  %``Quantized gauge theory on the fuzzy sphere as random matrix model,''
  Nucl.\ Phys.\ B {\bf 679}, 66 (2004)
  [arXiv:hep-th/0307075].
  %%CITATION = HEP-TH 0307075;%%

%
\bibitem{Karabali:2001te}
D.~Karabali, V.~P.~Nair and A.~P.~Polychronakos,
%``Spectrum of Schroedinger field in a noncommutative magnetic monopole,''
Nucl.\ Phys.\ B {\bf 627}, 565 (2002)
[arXiv:hep-th/0111249].
%%CITATION = HEP-TH 0111249;%%

%\cite{Carow-Watamura:2004ct}
\bibitem{Carow-Watamura:2004ct}
  U.~Carow-Watamura, H.~Steinacker and S.~Watamura,
  %``Monopole bundles over fuzzy complex projective spaces,''
  J.\ Geom.\ Phys.\  {\bf 54}, 373 (2005)
  [arXiv:hep-th/0404130].
  %%CITATION = HEP-TH 0404130;%%

\bibitem{GinspargWilson}P.~H.~Ginsparg and K.~G.~Wilson,
%``A Remnant Of Chiral Symmetry On The Lattice,''
Phys.\ Rev.\ D {\bf 25}, 2649 (1982).
%%CITATION = PHRVA,D25,2649;%% 

\bibitem{Neuberger}
H.~Neuberger,
%``Exactly massless quarks on the lattice,''
Phys.\ Lett.\ B {\bf 417}, 141 (1998)
[arXiv:hep-lat/9707022];
%%CITATION = HEP-LAT 9707022;%%
% 
%``Vector like gauge theories with almost massless fermions on the
%lattice,''
Phys.\ Rev.\ D {\bf 57}, 5417 (1998)
[arXiv:hep-lat/9710089];
%%CITATION = HEP-LAT 9710089;%%
% 
%``More about exactly massless quarks on the lattice,''
Phys.\ Lett.\ B {\bf 427}, 353 (1998)
[arXiv:hep-lat/9801031].
%%CITATION = HEP-LAT 9801031;%% 

\bibitem{Hasenfratzindex}
P.~Hasenfratz,
%``Prospects for perfect actions,''
Nucl.\ Phys.\ Proc.\ Suppl.\  {\bf 63}, 53 (1998)
[arXiv:hep-lat/9709110];
%%CITATION = HEP-LAT 9709110;%%
% 
P.~Hasenfratz, V.~Laliena and F.~Niedermayer,
%``The index theorem in QCD with a finite cut-off,''
Phys.\ Lett.\ B {\bf 427}, 125 (1998)
[arXiv:hep-lat/9801021].
%%CITATION = HEP-LAT 9801021;%% 

\bibitem{Luscher}M.~L\"uscher,
%``Exact chiral symmetry on the lattice and the Ginsparg-Wilson relation,''
Phys.\ Lett.\ B {\bf 428}, 342 (1998)
[arXiv:hep-lat/9802011].
%%CITATION = HEP-LAT 9802011;%% 

\bibitem{Nieder}F.~Niedermayer,
%``Exact chiral symmetry, topological charge and related topics,''
Nucl.\ Phys.\ Proc.\ Suppl.\  {\bf 73}, 105 (1999)
[arXiv:hep-lat/9810026].
%%CITATION = HEP-LAT 9810026;%% 

\bibitem{AIN2}
H.~Aoki, S.~Iso and K.~Nagao,
%``Ginsparg-Wilson relation, topological invariants and finite
%noncommutative geometry,''
Phys.\ Rev.\ D {\bf 67}, 085005 (2003)
[arXiv:hep-th/0209223].
%%CITATION = HEP-TH 0209223;%% 
%

%For a very short review, see
%%\bibitem{nagaolat03}
%K.~Nagao,
%%``Matrix model and Ginsparg-Wilson relation,''
%Nucl.\ Phys.\ Proc.\ Suppl.\  {\bf 129}, 501 (2004)
%[arXiv:hep-th/0309153].
%%%CITATION = HEP-TH 0309153;%% 

\bibitem{Madore}
J.~Madore,
%``The fuzzy sphere,''
Class.\ Quant.\ Grav.\  {\bf 9}, 69 (1992).
%%CITATION = CQGRD,9,69;%% 

%---------------------------------------------------------------
\bibitem{balagovi}
A.~P.~Balachandran, T.~R.~Govindarajan and B.~Ydri,
%``The fermion doubling problem and noncommutative geometry,''
Mod.\ Phys.\ Lett.\ A {\bf 15}, 1279 (2000)
[arXiv:hep-th/9911087];
%%CITATION = HEP-TH 9911087;%%
% 
%\bibitem{balaGW}
%A.~P.~Balachandran, T.~R.~Govindarajan and B.~Ydri,
%``The fermion doubling problem and noncommutative geometry. II,''
arXiv:hep-th/0006216.
%%CITATION = HEP-TH 0006216;%% 
 
%\cite{Ydri:2002nt}
\bibitem{Ydri:2002nt}
B.~Ydri,
%``Noncommutative chiral anomaly and the Dirac-Ginsparg-Wilson operator,''
JHEP {\bf 0308}, 046 (2003)
[arXiv:hep-th/0211209].
%%CITATION = HEP-TH 0211209;%% 


%\cite{Balachandran:2003ay}
\bibitem{Balachandran:2003ay}
A.~P.~Balachandran and G.~Immirzi,
%``The fuzzy Ginsparg-Wilson algebra:
%A solution of the fermion doubling problem,''
Phys.\ Rev.\ D {\bf 68}, 065023 (2003)
[arXiv:hep-th/0301242].
%%CITATION = HEP-TH 0301242;%% 

\bibitem{AIN3}
H.~Aoki, S.~Iso and K.~Nagao,
%``Ginsparg-Wilson relation and 't Hooft-Polyakov monopole on fuzzy 2-sphere,''
Nucl.\ Phys.\ B {\bf 684}, 162 (2004)
[arXiv:hep-th/0312199].
%%CITATION = HEP-TH 0312199;%% 
%


%------------------------------------------------------------
%
\bibitem{Nishimura:2001dq}
J.~Nishimura and M.~A.~Vazquez-Mozo,
%``Noncommutative chiral gauge theories on the lattice with manifest star-gauge invariance,''
JHEP {\bf 0108}, 033 (2001)
[arXiv:hep-th/0107110].
%%CITATION = HEP-TH 0107110;%% 



%
\bibitem{Iso:2002jc}
S.~Iso and K.~Nagao,
%``Chiral anomaly and Ginsparg-Wilson relation on the noncommutative torus,''
Prog.\ Theor.\ Phys.\  {\bf 109}, 1017 (2003)
[arXiv:hep-th/0212284].
%%CITATION = HEP-TH 0212284;%% 

%
\bibitem{Fujiwara:2002xh}
T.~Fujiwara, K.~Nagao and H.~Suzuki,
%``Axial anomaly with the overlap-Dirac operator in arbitrary dimensions,''
JHEP {\bf 0209}, 025 (2002)
[arXiv:hep-lat/0208057].
%%CITATION = HEP-LAT 0208057;%%


\bibitem{Nishimura:2002hw}
J.~Nishimura and M.~A.~Vazquez-Mozo,
%``Lattice perturbation theory in noncommutative geometry and 
%parity anomaly in 3D noncommutative QED,''
JHEP {\bf 0301}, 075 (2003)
[arXiv:hep-lat/0210017].
%%CITATION = HEP-LAT 0210017;%% 

%---------------------------------------------------------

\bibitem{AIMN}
H.~Aoki, S.~Iso, T.~Maeda and K.~Nagao,
  %``Dynamical generation of a nontrivial index on the fuzzy 2-sphere,''
  Phys.\ Rev.\ D {\bf 71}, 045017 (2005)
  %[Erratum-ibid.\ D {\bf 71}, 069905 (2005)]
  [arXiv:hep-th/0412052]

\bibitem{Myers:1999ps}
R.~C.~Myers,
%``Dielectric-branes,''
JHEP {\bf 9912} (1999) 022
[arXiv:hep-th/9910053].
%%CITATION = HEP-TH 9910053;%%

\bibitem{IKTW}
S.~Iso, Y.~Kimura, K.~Tanaka and K.~Wakatsuki,
%``Noncommutative gauge theory on fuzzy sphere from matrix model,''
Nucl.\ Phys.\ B {\bf 604}, 121 (2001)
[arXiv:hep-th/0101102].
%%CITATION = HEP-TH 0101102;%% 

\bibitem{Bal:2001cs}
S.~Bal and H.~Takata,
%``Interaction between two fuzzy spheres,''
Int.\ J.\ Mod.\ Phys.\ A {\bf 17}, 2445 (2002)
[arXiv:hep-th/0108002].
%%CITATION = HEP-TH 0108002;%%

\bibitem{Valtancoli:2002rx}
P.~Valtancoli,
%``Stability of the fuzzy sphere solution from matrix model,''
Int.\ J.\ Mod.\ Phys.\ A {\bf 18}, 967 (2003)
[arXiv:hep-th/0206075].
%%CITATION = HEP-TH 0206075;%%

\bibitem{Imai:2003vr}
T.~A.~Imai, Y.~Kitazawa, Y.~Takayama and D.~Tomino,
%``Quantum corrections on fuzzy sphere,''
Nucl.\ Phys.\ B {\bf 665}, 520 (2003)
[arXiv:hep-th/0303120]. 

\bibitem{Azuma:2004zq}
T.~Azuma, S.~Bal, K.~Nagao and J.~Nishimura,
%``Nonperturbative studies of fuzzy spheres in a matrix model with the
%Chern-Simons term,''
JHEP {\bf 0405}, 005 (2004)
[arXiv:hep-th/0401038];
%%CITATION = HEP-TH 0401038;%%
%
%\cite{Azuma:2004qe}
%\bibitem{Azuma:2004qe}
%  T.~Azuma, S.~Bal, K.~Nagao and J.~Nishimura,
   %``Dynamical aspects of the fuzzy CP(2) in the large N reduced model with a cubic term,''
  JHEP {\bf 0605}, 061 (2006)
  [arXiv:hep-th/0405277];
  %%CITATION = HEP-TH 0405277;%%
%
%\cite{Azuma:2004ie}
%\bibitem{Azuma:2004ie}
  T.~Azuma, K.~Nagao and J.~Nishimura,
  %``Perturbative dynamics of fuzzy spheres at large N,''
  JHEP {\bf 0506}, 081 (2005)
  [arXiv:hep-th/0410263].
  %%CITATION = HEP-TH 0410263;%%

%\cite{Castro-Villarreal:2004vh}
\bibitem{Castro-Villarreal:2004vh}
  P.~Castro-Villarreal, R.~Delgadillo-Blando and B.~Ydri,
   %``A gauge-invariant UV-IR mixing and the corresponding phase transition  for 
   %U(1) fields on the fuzzy sphere,''
  Nucl.\ Phys.\ B {\bf 704}, 111 (2005)
  [arXiv:hep-th/0405201].
  %%CITATION = HEP-TH 0405201;%%


\bibitem{ISTT}
%\bibitem{Ishiki:2006yr}
  G.~Ishiki, S.~Shimasaki, Y.~Takayama and A.~Tsuchiya,
  %``Embedding of theories with SU(2|4) symmetry into the plane wave matrix
  %model,''
  JHEP {\bf 0611}, 089 (2006)
  [arXiv:hep-th/0610038].
  %%CITATION = JHEPA,0611,089;%%

%\cite{Carow-Watamura:1996wg}
\bibitem{Carow-Watamura:1996wg}
U.~Carow-Watamura and S.~Watamura,
%``Chirality and Dirac operator on noncommutative sphere,''
Commun.\ Math.\ Phys.\  {\bf 183}, 365 (1997)
[arXiv:hep-th/9605003];
%%CITATION = HEP-TH 9605003;%%
%
%\cite{Carow-Watamura:1998jn}
%\bibitem{Carow-Watamura:1998jn}
%U.~Carow-Watamura and S.~Watamura,
%``Noncommutative geometry and gauge theory on fuzzy sphere,''
Commun.\ Math.\ Phys.\  {\bf 212}, 395 (2000)
[arXiv:hep-th/9801195].
%%CITATION = HEP-TH 9801195;%%
 
 
%------D_GKP-----------------------------------------
%\cite{Grosse:1994ed}
\bibitem{Grosse:1994ed}
H.~Grosse and J.~Madore,
%``A Noncommutative version of the Schwinger model,''
Phys.\ Lett.\ B {\bf 283}, 218 (1992);
%%CITATION = PHLTA,B283,218;%%
 
H.~Grosse and P.~Presnajder,
%``The Dirac operator on the fuzzy sphere,''
Lett.\ Math.\ Phys.\  {\bf 33}, 171 (1995);
%%CITATION = LMPHD,33,171;%%
 
H.~Grosse, C.~Klimcik and P.~Presnajder,
%``Field theory on a supersymmetric lattice,''
Commun.\ Math.\ Phys.\  {\bf 185}, 155 (1997)
[arXiv:hep-th/9507074];
%%CITATION = HEP-TH 9507074;%%
%
%``N = 2 superalgebra and non-commutative geometry,''
arXiv:hep-th/9603071.
%%CITATION = HEP-TH 9603071;%%
%---------------------------------------------------------

 
 
\bibitem{chiral_anomaly}
H.~Grosse and P.~Presnajder,
%``A treatment of the Schwinger model within noncommutative geometry,''
arXiv:hep-th/9805085;
%%CITATION = HEP-TH 9805085;%%
%
%``A Noncommutative Regularization Of The Schwinger Model,''
Lett.\ Math.\ Phys.\  {\bf 46}, 61 (1998). 
%%CITATION = LMPHD,46,61;%%
 
\bibitem{chiral_anomaly2}
P.~Presnajder,
%``The origin of chiral anomaly and the noncommutative geometry,''
J.\ Math.\ Phys.\  {\bf 41}, 2789 (2000)
[arXiv:hep-th/9912050].
%%CITATION = HEP-TH 9912050;%% 
 
\bibitem{AIN1}
H.~Aoki, S.~Iso and K.~Nagao,
%``Chiral anomaly on fuzzy 2-sphere,''
Phys.\ Rev.\ D {\bf 67}, 065018 (2003)
[arXiv:hep-th/0209137].
%%CITATION = HEP-TH 0209137;%%


\bibitem{'tHooft:1974qc}
  G.~'t Hooft,
  %``MAGNETIC MONOPOLES IN UNIFIED GAUGE THEORIES,''
  Nucl.\ Phys.\ B {\bf 79}, 276 (1974).
  %%CITATION = NUPHA,B79,276;%%

%\cite{Luscher:1981zq}
\bibitem{Luscher:1981zq}
  M.~Luscher,
  %``Topology Of Lattice Gauge Fields,''
  Commun.\ Math.\ Phys.\  {\bf 85}, 39 (1982).
  %%CITATION = CMPHA,85,39;%%

\bibitem{Hernandez:1998et}
  P.~Hernandez, K.~Jansen and M.~Luscher,
  %``Locality properties of Neuberger's lattice Dirac operator,''
  Nucl.\ Phys.\  B {\bf 552}, 363 (1999)
  [arXiv:hep-lat/9808010].
  %%CITATION = NUPHA,B552,363;%%

\bibitem{Luscher:1998kn}
  M.~Luscher,
  %``Topology and the axial anomaly in abelian lattice gauge theories,''
  Nucl.\ Phys.\ B {\bf 538}, 515 (1999)
  [arXiv:hep-lat/9808021];
  %%CITATION = HEP-LAT 9808021;%%
%
%\bibitem{Luscher:1998du}
 % M.~Luscher,
  %``Abelian chiral gauge theories on the lattice with exact gauge
  %invariance,''
  Nucl.\ Phys.\ B {\bf 549}, 295 (1999)
  [arXiv:hep-lat/9811032].
  %%CITATION = HEP-LAT 9811032;%%

\bibitem{aokiiso}
H.Aoki and S.Iso,
in preparation


%\cite{Fujikawa:1999ku}
\bibitem{Fujikawa:1999ku}
  K.~Fujikawa,
  %``Relation Tr(gamma(5)) = 0 and the index theorem in lattice gauge  theory,''
  Phys.\ Rev.\ D {\bf 60}, 074505 (1999)
  [arXiv:hep-lat/9904007].
  %%CITATION = HEP-LAT 9904007;%%



%%%%%%%%%%%%%%%%%%%%%%%%%%%%%%%%%%%%%%%%%%%%%%%%%%%%%%%%%
%\cite{Nagao:2005st}
\bibitem{Nagao:2005st}
  K.~Nagao,
  %``Admissibility condition and nontrivial indices on a noncommutative torus,''
  Phys.\ Rev.\ D {\bf 73}, 065002 (2006)
  [arXiv:hep-th/0509034].
  %%CITATION = HEP-TH 0509034;%%

\bibitem{ANS}
%\cite{Aoki:2006sb}
%\bibitem{Aoki:2006sb}
  H.~Aoki, J.~Nishimura and Y.~Susaki,
  %``The index theorem in gauge theory on a discretized 2d non-commutative torus,''
  JHEP {\bf 0702}, 033 (2007)  [arXiv:hep-th/0602078];
  %%CITATION = HEP-TH 0602078;%%
%\cite{Aoki:2006zi}
%\bibitem{Aoki:2006zi}
  %H.~Aoki, J.~Nishimura and Y.~Susaki,
  %``Suppression of topologically nontrivial sectors in gauge theory on 2d non-commutative geometry,''
  arXiv:hep-th/0604093.
  %%CITATION = HEP-TH 0604093;%%

%\bibitem{AIM} work in progress.

%\cite{Wu:1976ge}
\bibitem{Wu:1976ge}
  T.~T.~Wu and C.~N.~Yang,
  %``Dirac Monopole Without Strings: Monopole Harmonics,''
  Nucl.\ Phys.\ B {\bf 107}, 365 (1976).
  %%CITATION = NUPHA,B107,365;%%

%\cite{Rubakov:1982fp}
\bibitem{Rubakov:1982fp}
  V.~A.~Rubakov,
  %``Adler-Bell-Jackiw Anomaly And Fermion Number Breaking In The Presence Of A
  %Magnetic Monopole,''
  Nucl.\ Phys.\ B {\bf 203}, 311 (1982).
  %%CITATION = NUPHA,B203,311;%%

\end{thebibliography}
\end{document}